\documentclass[twocolumn,preprintnumbers,amsmath,amssymb,nofootinbib]{revtex4}
\pdfoutput=1

\usepackage{graphicx}% Include figure files
\usepackage{dcolumn}% Align table columns on decimal point
\usepackage{bm}% bold math

\newcommand{\eps}{\varepsilon}

\def\<{\langle} \def\>{\rangle}

\usepackage{amsmath,amssymb,amscd,enumerate,mathrsfs,latexsym}
\usepackage{graphicx}

\def\RR{\mathbb{R}}

\def\<{\langle} \def\>{\rangle}

\newcommand{\uu}[1]{{\boldsymbol #1}}

\def\ff{\uu{f}}
\def\ttheta{\uu{\theta}}
\def\xx{\uu{x}}
\def\zz{\uu{z}}
\def\eeta{\uu{\eta}}

\begin{document}

%%%%%%%%%%%%%%%%%%%%%%%%%%%%%%

\title{Single-Sweep Methods for Free Energy Calculations}

%% Enter authors via the \author command.  Use \thanks to indicate the
%% corresponding author.  Use \affil to define affiliations.

%% \author{<author name>
%% for corresponding author only, enter text like this:
%% \thanks{<To whom correspondence should be addressed. E-mail:
%% <authoremail>}%
%% \thanks will print at the bottom of the page. For the corresponding author:
%% \thanks{To whom correspondence should be addressed. Email: <author email>}
%% \affil{<number>}{<Institution>}} One number for each institution.
%% The same number should be used for authors that
%% are affiliated with the same institution, after the first time
%% only the number is needed, ie, \affil{number}{text}, \affil{number}{}
%% Then, before last author ...
%% \and
%% \author{<author name>
%% \affil{<number>}{}}

%% For example,
%% \author{Diego C\'ordoba\affil{1}{Consejo Superior de
%% Investigaciones Cient\'if\,icas c/ Serrano 123, 28006 Madrid, Spain},
%% Charles Fefferman\thanks{To whom correspondence should be addressed. E-mail:
%% cfz@math.princeton.edu}%
%% \affil{2}{Princeton University  Department of Mathematics,
%% Washington Road, Princeton, NJ 08544-1000}, \and
%% Jos\'e Luis Rodrigo\affil{2}{}}

\author{Luca Maragliano}%
 \email{maraglia@cims.nyu.edu}
\affiliation{%
  Courant Institute of Mathematical Sciences, New York University, New
  York, NY 10012, USA
}%

\author{Eric Vanden-Eijnden}%
 \email{eve2@cims.nyu.edu}
\affiliation{%
  Courant Institute of Mathematical Sciences, New York University, New
  York, NY 10012, USA
}%

%\date{}

%%%%%%%%%%%%%%%%%%%%%%%%%%%%%%%%%%%%%%%%%%%%%%%%%%%%%%%%%%%%%%%%

\begin{abstract}
  A simple, efficient, and accurate method is proposed to map
  multi-dimensional free energy landscapes. The method combines the
  temperature-accelerated molecular dynamics (TAMD) proposed in
  [Maragliano \& Vanden-Eijnden, Chem. Phys. Lett.  \textbf{426}, 168
  (2006)] with a variational reconstruction method using radial-basis
  functions for the representation of the free energy.  TAMD is used
  to rapidly sweep through the important regions of the free energy
  landscape and compute the gradient of the free energy locally at
  points in these regions. The variational method is then used to
  reconstruct the free energy globally from the mean force at these
  points. The algorithmic aspects of the single-sweep method are explained
  in detail, and the method is tested on simple examples, compared to
  metadynamics, and finally used to compute the free energy of the
  solvated alanine dipeptide in two and four dihedral angles.
\end{abstract}

%% When adding keywords, separate each term with a straight line: |

\maketitle

\section{Introduction}
\label{sec:intro}

%\vspace{1.cm}

The free energy (or potential of mean force) is the thermodynamic
force driving structural processes such as conformational changes of
macromolecules in aqueous solution, ligand binding at the active site
of an enzyme, protein-protein association, etc. The free energy gives
information about both the rate at which these processes occur and the
mechanism by which they occur.  This makes free energy calculations a
central issue in biophysics.  Molecular dynamics (MD) simulations
provide a tool for performing such calculations on a computer in a way
which is potentially both precise and inexpensive
(e.g.~\cite{fe1,fe2,fe3}). Since a free energy is in essence the
logarithm of a probability density function (see~(\ref{eq:free}) below
for a precise definition) it can in principle be calculated by
histogram methods based on the binning of an MD trajectory. This
direct approach, however, turns out to be unpractical in general
because the time scale required for the trajectory to explore all the
relevant regions of configuration space is prohibitively long.
Probably the best known and most widely used technique to get around
this difficulty is the weighted histogram analysis method
(WHAM)~\cite{wham}. Following~\cite{valleau}, WHAM adds artificial
biasing potentials to maintain the MD system in certain umbrella
sampling windows. WHAM then recombines in an optimal way the
histograms from all the biased simulations to compute the free energy.
WHAM is much more efficient than the direct sampling approach, and
generalizations such as~\cite{karplus} alleviate somewhat the problem
of where to put the umbrella windows (usually, this requires some
\textit{a~priori} knowledge of the free energy landscape). In
practice, however, WHAM remains computationally demanding and it only
works to compute the free energy in 2 or 3 variables. An interesting
alternative to WHAM is metadynamics~\cite{Laio02,Laio03}.  In essence
metadynamics is a way to use an MD trajectory to place inverted
umbrella sampling windows on-the-fly and use these windows both to
bias the MD simulation and as histogram bins to sample the free energy
directly (thereby bypassing the need of further histogram analysis in
each window).

Both WHAM and metadynamics compute the free energy directly by
histogram methods, but an alternative approach is possible. Unlike the
free energy which is a global quantity, its negative gradient (known
as the mean force) can be expressed in terms of a local expectation
and thereby computed at a given point in the free energy landscape.
This is the essence of the blue moon sampling strategy~\cite{BlueM99}
and it offers the possibility to calculate first the mean force at a
given set of locations, then use this information to reconstruct the
free energy globally. In one dimension, this approach is known as
thermodynamic integration and it goes back to
Kirkwood~\cite{kirkwood}. In higher dimensions, however, this way to
compute free energies has been impeded by two issues. The first is
where to place the points at which to compute the mean force, and the
second is how to reconstruct the free energy from these data

In this paper, we propose a method, termed single-sweep method, which
addresses both of these issues in two complementary but independent
steps.  In a first step, we use the temperature-accelerated molecular
dynamics (TAMD) proposed in~\cite{tamd} (see
also~\cite{varo02,Tuck01}) to quickly sweep through the important
regions of the free energy landscape and identify points in these
regions where to compute the mean force.  In the second step we then
reconstruct the free energy globally from the mean force by
representing the free energy using radial-basis functions, and
adjusting the parameters in this representation via minimization of an
objective function.  

The single-sweep method is easy to use and
implement, does not require \textit{a~priori} knowledge of the free
energy landscape, and can be applied to map free energies in several
variables (up to four, as demonstrated here, and probably more). The
single-sweep method is also very efficient, especially since the mean
force calculations can be performed using independent calculations on
distributed processors (i.e. using grid computing facilities~\cite{grid1,grid2}).

The reminder of this paper is organized as follows. In
Sec.~\ref{sec:theory}, we describe the two steps of the single-sweep
method in detail, starting with the second one for convenience. In
Sec.~\ref{sec:app1} we illustrate the method on a simple two-dimensional
example. This example is then used for comparison with metadynamics
in Sec.~\ref{sec:metadynamics}.  In Sec.~\ref{sec:AD} we use the
single-sweep method to compute the free energy of alanine dipeptide
(AD) in solution in two and in four of its dihedral angles.  Finally,
concluding remarks are made in Sec.~\ref{sec:conclu} and the details
of the MD calculation on AD are given in Appendix~\ref{sec:MD}

\section{The Single-Sweep method}
\label{sec:theory}

\subsection{Free energy representation and reconstruction}
\label{sec:reconstruct}

We shall consider a molecular system with $n$ degrees of freedom whose
position in configuration space~$\Omega \subseteq \RR^n$ will be
denoted by $\xx$.  We also introduce a set of $N$ collective variables
$\ttheta(\xx) = (\theta_1(\xx), \ldots, \theta_N(\xx))$ which are
functions of $\xx$ such as torsion angles, interatomic distances, etc.
If $V(\xx)$ denotes the potential energy of the system and $1/\beta$
its temperature, the free energy $A(\zz)$ in the variables
$\ttheta(\xx)$ is defined as
\begin{equation}
  \label{eq:free}
  A(\zz) = - \beta^{-1} \log \int_{\Omega} e^{-\beta V(\xx)} 
  \delta(\ttheta(\xx) - \zz) dx
\end{equation}
so that $e^{-\beta A(\zz)}$ is, up to a proportionality constant, the
probability density function (PDF) of the variables~$\ttheta(\xx)$.

As mentioned in the introduction, the negative gradient of the free
energy, $\ff(\zz) = -\nabla_{\!z} A(\zz)$, is known as the mean force,
and it can be computed locally at point~$\zz$ via calculation of an
expectation (see~(\ref{eq:meanforceapprox}) below). In this section,
we shall suppose that we have obtained an estimate of $\ff(\zz)$ at
points $\zz_1, \ldots, \zz_K$, and we focus on the reconstruction of
the free energy $A(\zz)$ from these data. A specific way to pick these
points and compute $\ff_{\!1}\approx \ff(\zz_1), \ldots,
\ff_{\!K}\approx \ff(\zz_K)$ will be given in
Sec.~\ref{sec:TAMD}, but it is worth pointing out that the
reconstruction method proposed here works with data set collected in
any other ways.

Our reconstruction method uses a radial-basis function representation
for the free energy~$A(\zz)$ with centers at $\zz_1, \ldots,
\zz_K$~\cite{radialbasis0,radialbasis1}:
\begin{equation}
  \label{eq:radialbasis}
  \tilde A(\zz) = \sum_{k=1}^K a_k \varphi_\sigma(|\zz- \zz_k|) +C.
\end{equation}
Here $C$ is a constant used to adjust the overall height of $\tilde
A(\zz)$ but is otherwise irrelevant, $|\cdot|$ denotes the Euclidean
norm in~$\RR^N$, and $\varphi_\sigma(u) = \varphi(u/\sigma) $ where
$\varphi(u)$ is a radial-basis function; a convenient choice is to use
the Gaussian packet
\begin{equation}
  \label{eq:kernel}
  \varphi(u) = e^{- \frac12u^2}
\end{equation}
though other radial-basis functions (multiquadric, Sobolev splines,
Wendland, etc.~\cite{radialbasis0}) can be used as well, see
Sec.~\ref{sec:AD}.  In (\ref{eq:radialbasis}) the heights $a_k$ and
the radial-basis function width $\sigma>0$ are adjustable parameters
which we determine by minimizing over $a_k$ and $\sigma$ the following
objective function, which measures the discrepancy between the
negative gradient of the function $\tilde A(z)$ in
(\ref{eq:radialbasis}) at the centers $\zz_k$, $\nabla_z \tilde
A(\zz_k) = \sum_{k'=1}^K a_{k'} \nabla_z\varphi_\sigma(|\zz_k-
\zz_{k'}|)$, and the mean force $\ff_{\!k}$ estimated at these
centers:
\begin{equation}
  \label{eq:objective}
    E(a,\sigma) = \sum_{k=1}^K \Big| \sum_{k'=1}^K a_{k'}
    \nabla_z\varphi_\sigma(|\zz_k- \zz_{k'}|) + \ff_{\!k}\Big|^2.
\end{equation}
Before explaining how we perform this minimization, let us give
several reasons why the radial-basis
representation~(\ref{eq:radialbasis}) for $A(\zz)$ is natural and
convenient.  First, the centers $\zz_k$ in~(\ref{eq:radialbasis}) do
not have to lie on a regular grid, which permits to use mean force data
collected anywhere. Second, the representation~(\ref{eq:radialbasis})
can be used in any dimension. Third, this representation has very good
convergence properties, i.e.  a small number of centers gives an
accurate representation of~$A(\zz)$. In fact, unlike standard
representations based e.g.  on linear interpolation on a regular grid,
the rate of convergence in $K$ of the representation
in~(\ref{eq:radialbasis}) can be made independent of $N$ (a feature
which the radial-basis representations share with sparse
grids~\cite{radialbasis2,sparsegrid}).

Going back to the minimization of $E(a,\sigma)$, it can be performed
as follows. For fixed $\sigma$, the~$a^\star_k$ minimizing
(\ref{eq:objective}) solve the following linear algebraic system
\begin{equation}
  \label{eq:linalg}
  \sum_{k'=1}^K B_{k,k'}(\sigma) a^\star_{k'}(\sigma) = c_k(\sigma)
\end{equation}
where $B_{k,k'}(\sigma)$ and $c_k(\sigma)$ are given by
\begin{equation}
  \label{eq:Bc}
  \begin{aligned}
    B_{k,k'}(\sigma) &= \sum_{k''=1}^K \nabla_z\varphi_\sigma(|\zz_k-
    \zz_{k''}|)\cdot \nabla_z\varphi_\sigma(|\zz_{k''}- \zz_{k'}|),\\
    c_k(\sigma) &= -\sum_{k'=1}^K \nabla_z\varphi_\sigma(|\zz_k-
    \zz_{k'}|)\cdot \ff_{\!k'} \, .
  \end{aligned} 
\end{equation}
Given the centers $\zz_k$ and the estimates $\ff_k$ of the mean force at
these centers, the coefficients $B_{k,k'}$ and $c_k$ can be easily
computed, and the linear system (\ref{eq:linalg}) can be solved by any
standard technique, e.g. Gaussian elimination.  Once the
solution~$a^\star_k\equiv a^\star_k(\sigma)$ of~(\ref{eq:linalg}) is
determined, to find the optimal $\sigma^\star$ satisfying
$E(a^\star(\sigma^\star),\sigma^\star) = \min_\sigma
E(a^\star(\sigma),\sigma)$ we compute the residual
$E(a^\star(\sigma),\sigma)$ for increasing values of $\sigma$ starting
from the distance between the centers. 
More sophisticated procedures could be used to
minimize $E(a^\star(\sigma),\sigma)$ over $\sigma$, but
the brute force method that we used proved to be efficient enough
because computing successive solutions of (\ref{eq:linalg}) for
various~$\sigma$ is very fast.  To measure the error in the
approximation, we used the residual per center defined as
\begin{equation}
  \label{eq:resid}
  e_2(\sigma) = E^{1/2}(a^\star(\sigma),\sigma)/K \, ,
\end{equation}
which reaches its minimum value at the same $\sigma^\star$ as
$E(a^\star(\sigma),\sigma)$.

Overall, the procedure is simple and inexpensive since the
determination of $a^\star_k$ at fixed~$\sigma$ is computationally
straightforward and cheap, and can be easily repeated to perform the
one-dimensional minimization over~$\sigma$.  One caveat that we should
mention, however, is that the condition number of the matrix
$B_{k,k'}(\sigma)$ increases rapidly when the number of centers and/or
$\sigma$ increase.  This is a known problem of radial-basis
functions~\cite{radialbasis1}.  To avoid any problems, we capped the
admissible condition number at $10^{12}$ and, in situations where this
threshold value was reached while $e_2(\sigma)$ was still decreasing,
picked for $\sigma^\star$ the corresponding value of $\sigma$.  These
situations only occurred in the two-dimensional example (see
Sec.~\ref{sec:app1}) when a lot of centers were used (500 or more, i.e.
much more than what will be used in the AD example), and even in these
cases, such a large condition number did not lead to any noticeable
loss of accuracy in the results (even though the coefficients
$a^\star_k$ were then very large).  We also observed that, given a
number of centers~$\zz_k$ and a value of~$\sigma$, the condition
number is typically lower when the dimension of $\zz$ is larger.
Finally, we observed that the condition number was much lower with the
Wendland radial-basis function (see (\ref{eq:wend}) in
Sec.~\ref{sec:AD}) than with the Gaussian radial-basis
function~(\ref{eq:kernel}).

\subsection{TAMD for sweeping}
\label{sec:TAMD}

It remains to explain how to identify the centers $\zz_1,\ldots \zz_K$
and estimate the mean force at these points.  Following~\cite{tamd},
we will do so using the extended system
\begin{equation}
  \label{eq:tamd}
  \begin{cases}
    \begin{aligned}
      M\ddot \xx &=-\nabla_x V(\xx) -\kappa \sum_{\alpha=1}^N
      (\theta_\alpha(\xx) - z_\alpha)\nabla_x
      \theta_\alpha(\xx)\\[-4pt] & \quad + \text{thermostat terms at
        $\beta^{-1}$}
    \end{aligned}
    \\[20pt]
    \displaystyle\gamma \dot \zz = \kappa (\ttheta(\xx) - \zz) + \sqrt{2 \gamma
      \bar \beta^{-1}}\, \eeta(t)
  \end{cases}
\end{equation}
where $M$ is the mass matrix, $\eeta(t)$ is a white-noise, i.e. a
Gaussian process with mean 0 and covariance $\<\eta_{\alpha}(t)
\eta_{\alpha'} (t')\> = \delta_{\alpha\alpha'} \delta(t-t')$, and
$\kappa>0$, the friction coefficient $\gamma>0$ and the artificial
inverse temperature $1/\bar\beta\ (\not=1/\beta)$ are parameters whose
role we explain now. 

The system in~(\ref{eq:tamd}) describes the motion of $\xx$ and $\zz$
over the extended potential
\begin{equation}
  \label{eq:pote}
  U_\kappa(\xx,\zz) = V(\xx) + \tfrac12 \kappa |\ttheta(\xx) - \zz|^2.
\end{equation}
As shown in~\cite{tamd}, by adjusting the parameter $\kappa$ so that
$\zz(t) \approx \ttheta(\xx(t))$ and the friction coefficient $\gamma$
so that $\zz$ moves slower than $\xx$, one can generate a trajectory
$\zz(t)$ in $z$-space which effectively moves at the artificial
temperature $1/\bar\beta$ on the free energy computed at the physical
temperature $1/\beta$.  By taking $1/\bar\beta > 1/\beta$, the
$\zz(t)$ trajectory visits rapidly the regions where the free energy
is relatively low (i.e.  within a range of a few~$1/\bar\beta$) even
if these regions are separated by barriers which the system would take
a long time to cross at the physical temperature $1/\beta$. This gives
us a way to determine automatically where are the relevant regions in
free energy space.

In~\cite{tamd}, the extended system in~(\ref{eq:tamd}) was proposed to
sample the free energy landscape directly.  Here, we make a different
use of~(\ref{eq:tamd}): we utilize the trajectory~$\zz(t)$ to rapidly
sweep through $\zz$-space and generate the centers $\zz_1, \ldots,
\zz_K$ used in the radial-basis representation~(\ref{eq:radialbasis}).
Specifically, we start from $\zz(0)= \zz_1$, then deposit a new center
$\zz_k$ along $\zz(t)$ each time $\zz(t)$ reaches a point which is
more than a prescribed distance~$d$ away from all the previous
centers, where $d$ is a parameter controlling the density of the
covering by the centers (the smaller $d$, the higher the number of
centers deposited).  At the same time, at each of these
centers~$\zz_k$, we launch a simulation of~(\ref{eq:tamd}) with
$\zz(t)=\zz_k$ fixed, i.e we use
\begin{equation}
  \label{eq:tamd2}
    \begin{aligned}
      M\ddot \xx &=-\nabla_x V(\xx) -\bar \kappa \sum_{\alpha=1}^N
      (\theta_\alpha(\xx) - z_{k,\alpha})\nabla_x \theta_\alpha(\xx)\\[-4pt] 
      & \quad + \text{thermostat terms at $\beta^{-1}$}
    \end{aligned}
\end{equation}
and compute:
\begin{equation}
  \label{eq:meanforceapprox}
  \ff_{\!k} = \frac1T \int_0^T \bar\kappa 
  \left(\zz_k -\ttheta(\xx(t))\right) dt  \, .
\end{equation}
The calculations of these time averages are independent of each other,
and hence they can be distributed, using (ideally) at least one
processor per center $\zz_k$, an approach that optimally fits
with the purposes of grid computing~\cite{grid1,grid2}. 
The estimator in~(\ref{eq:meanforceapprox})
has the advantage of being simple, but it introduces an error due to
the finiteness of~$\bar\kappa$.  This error can be decreased by
using~$\bar\kappa$ in~(\ref{eq:tamd2}) and~(\ref{eq:meanforceapprox})
larger than $\kappa$ in~(\ref{eq:tamd}), or even eliminated by using
constrained instead of restrained simulations and using the blue-moon
estimator for the mean force~\cite{chemphyschem,free_cpam}.

Once the centers $\zz_1, \ldots, \zz_k$ have been deposited and the
estimates $\ff_{\!1},\ldots, \ff_{\!K}$ of the mean force at these
centers have been obtained, we use the reconstruction procedure
explained in Sec.\ref{sec:reconstruct} and compute the optimal set of
coefficients $a^\star_k$ and the optimal $\sigma^\star$ to use in the
representation~(\ref{eq:radialbasis}) for~$A(\zz)$.

\medskip

We conclude this section by stressing that using the extended
dynamics~(\ref{eq:tamd}) to sweep through $\zz$-space and deposit the
centers~$\zz_k$ is very different than using it to sample $A(\zz)$
directly, which makes our approach very different from WHAM or
metadynamics. Unlike with sampling, revisiting twice a region in
$\zz$-space is unnecessary and even undesirable since no new center
will be deposited. The accuracy of the reconstruction depends on the
number of centers and the accuracy at which the mean force is computed
in~(\ref{eq:meanforceapprox}) much more than the precise locations
where the centers are deposited.  An important practical consequence
is that it is rather straightforward to pick the parameters $\kappa$
and $\gamma$ in~(\ref{eq:tamd}) since the final result is robust
against variations in these parameters.

\begin{figure}[t]
  \centerline{\includegraphics[width=3.25in]{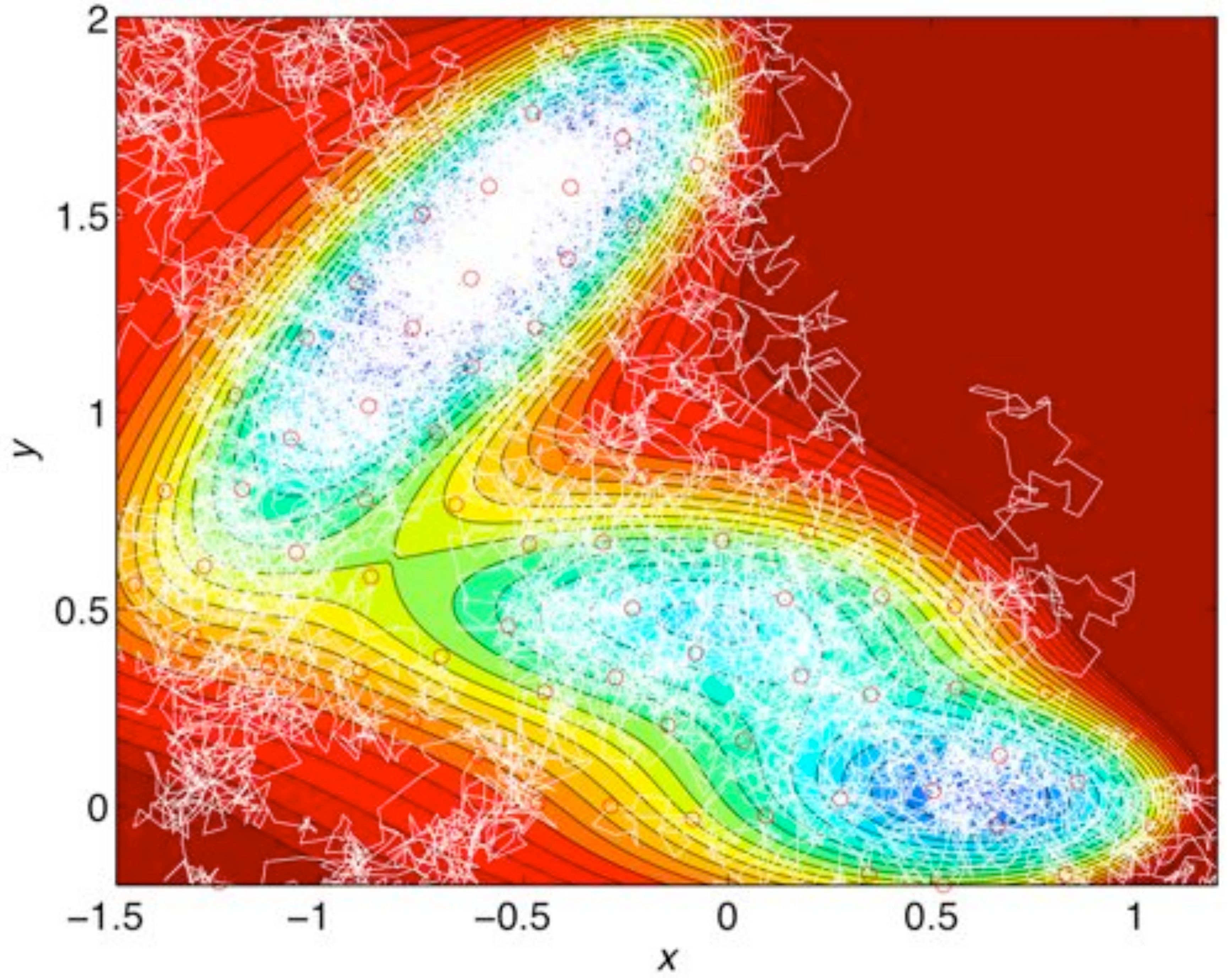}}
  \caption{%
    A trajectory generated by simulating~(\ref{eq:tamdex}) by forward
    Euler with a time-step ${\mathit\Delta }t = 2\cdot10^{-5}$ for
    $2\cdot10^4$ steps (white curve) shown above the contour plot of
    the Mueller potential (with $29$ level sets evenly distributed
    between $V=0$ and $V=180$ in a scale where the minimum of the
    potential is $V=0$).  The red circles are the locations of the
    centers deposited along the trajectory using $d=0.175$. In this
    run, $174$ centers were deposited.}
  \label{fig:1}
\end{figure}

\section{Two-dimensional illustrative example}
\label{sec:app1}

\begin{figure}[t]
  \centerline{\includegraphics[width=3.25in]{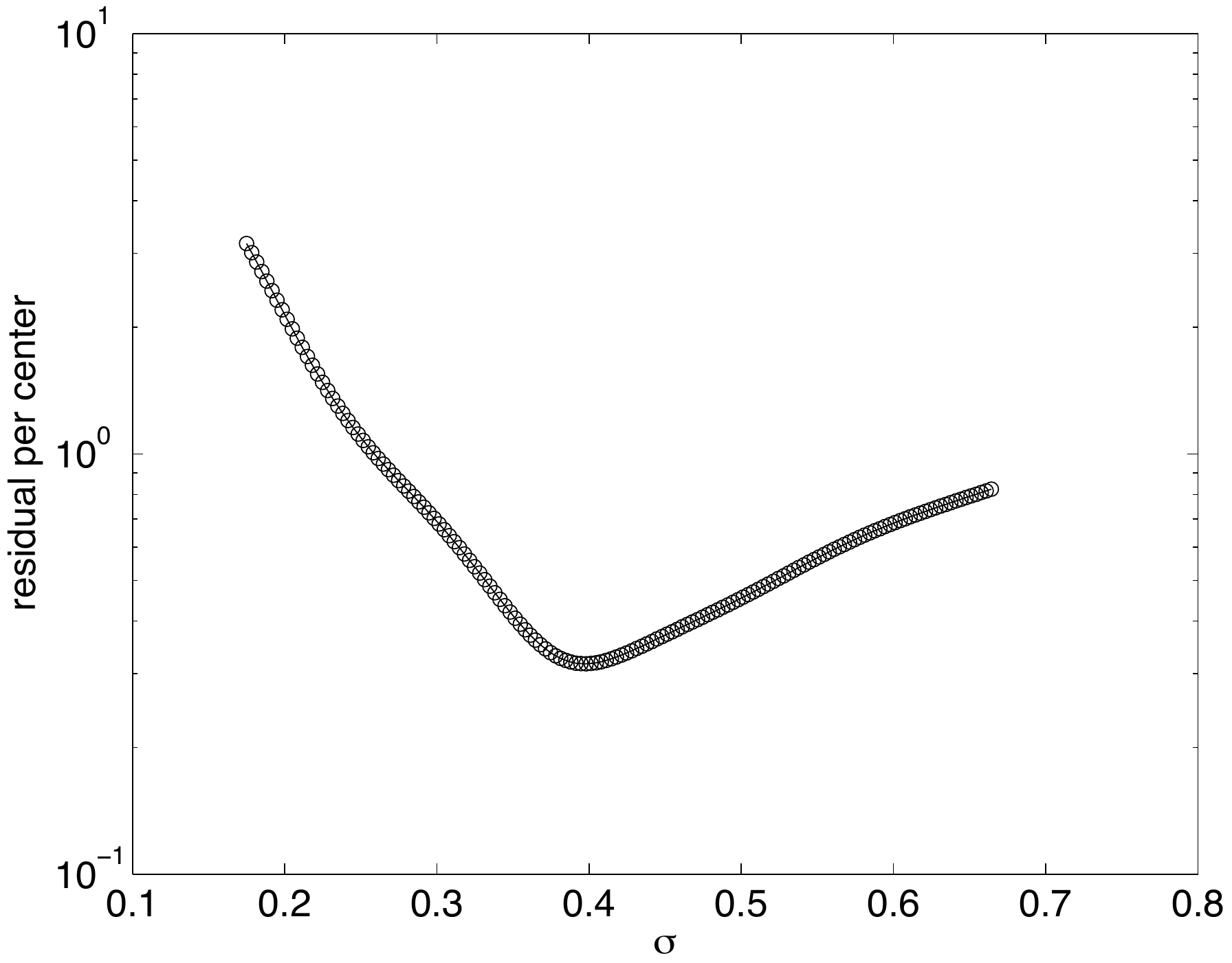}} 
    \caption{% 
      Residual per center~$e_2(\sigma)$ defined in~(\ref{eq:resid})
      for the reconstruction of the Mueller potential with the $2\cdot
      10^4$ steps single-sweep trajectory shown in Fig.~\ref{fig:1}.
      The optimal $\sigma$ for this run was $\sigma^\star = 0.398$.}
      \label{fig:muelsweepres}
\end{figure}

\begin{figure}[tbp]
  \centerline{\includegraphics[width=3.25in]{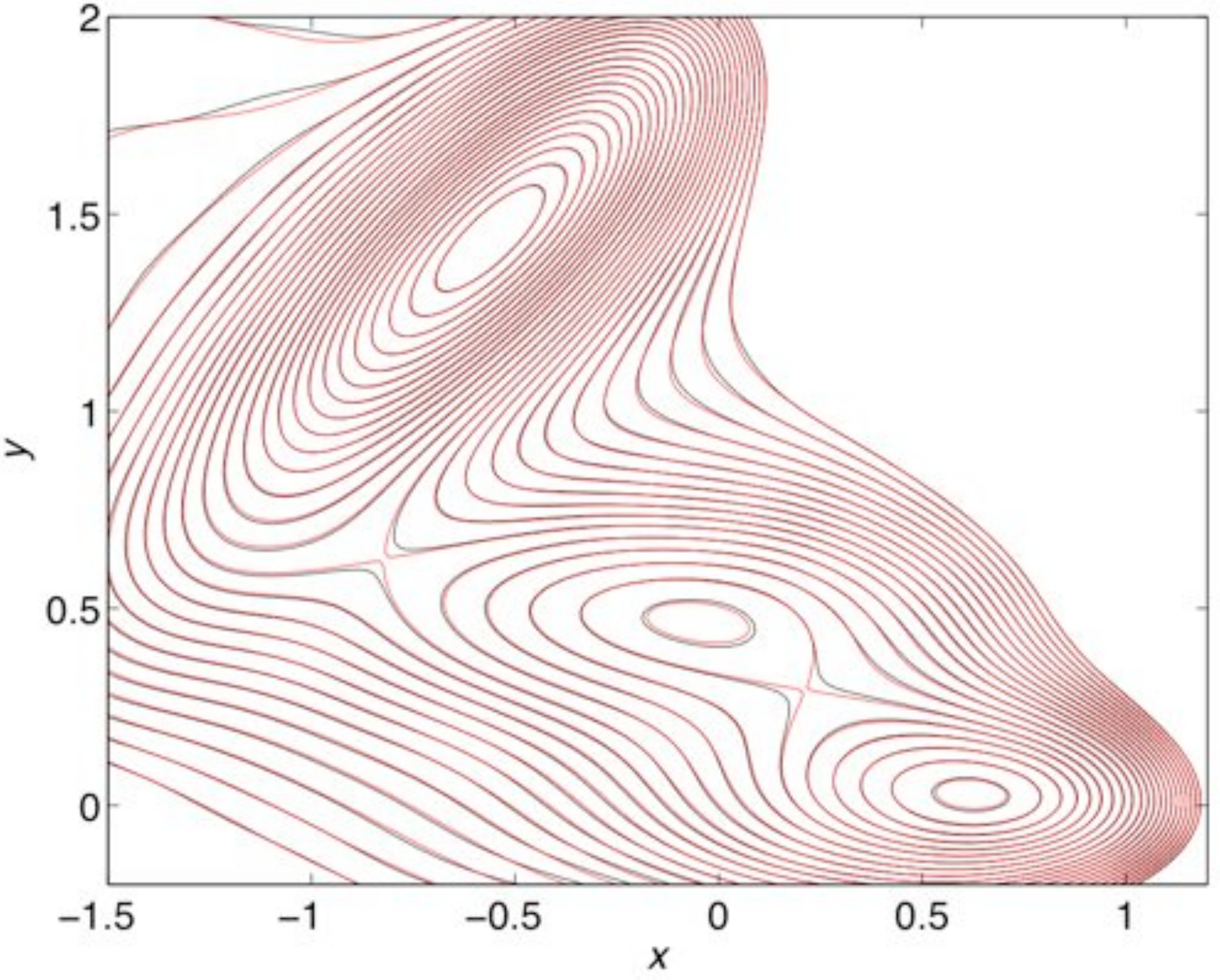}}
  \caption{%
    Comparison between the level sets of the original Mueller
    potential (red curves) and the reconstructed potential
    using~(\ref{eq:radialbasis}) (black curve). Here we use the $174$
    centers shown in Fig.~\ref{fig:1}. The optimal $\sigma$ is
    $\sigma^\star = 0.398$, see Fig.~\ref{fig:muelsweepres}.  We show
    29 level sets evenly distributed between $V=0$ and $V=180$.  The
    level sets of the reconstructed potential and the original one are
    in so close agreement that they can only be distinguished in some
    localized regions (e.g. near the saddle point between the two
    minima in the lower right corner). }
  \label{fig:2}
\end{figure}

Since, given the location of the centers $\zz_k$, the mean force
estimation at these points is quite standard, as a first illustration
we use a two-dimensional example for which $\ttheta(\xx) \equiv \xx =
(x,y)$ and $A(\zz) \equiv V(\xx) $ where $V(\xx)$ is the Mueller
potential~\cite{mueller}.  In this case, there is no need to extend
the system as in~(\ref{eq:tamd}), and the temperature accelerated
dynamics simply reduces to (setting $\gamma=1$ by appropriate
rescaling of time)
\begin{equation}
  \label{eq:tamdex}
    \dot \xx = -\nabla V(\xx) + \textstyle\sqrt{2\bar\beta^{-1}} \, 
    \eeta(t) \, .
\end{equation}
Fig.~\ref{fig:1} shows a TAMD trajectory generated by
solving~(\ref{eq:tamdex}) by forward Euler with the initial condition
$(x(0),y(0))=(1,0)$ and a time-step of $\mathit{\Delta } t =
2\cdot10^{-5}$ for $2\cdot10^4$ time-steps at $1/\bar \beta= 40$ (for
comparison the energy barrier between the two main minima of the
Mueller potential is about $100$). Also shown are the centers
$\zz_k\equiv (x_k,y_k)$ obtained by depositing a new center along the
trajectory each time the trajectory reaches a point which is $d=0.175$
away for all the previous centers. In this run, $174$ centers were
deposited.  At the centers, we used $-\nabla V(x_k,y_k) = \ff_{\!k}$
as estimate of the ``mean force'' (i.e.  there is no sampling error in
the present example).  We then used these data to reconstruct the free
energy as explained in Sec.~\ref{sec:reconstruct}.
Fig.~\ref{fig:muelsweepres} shows the residual per center
$e_2(\sigma)$ defined in~(\ref{eq:resid}). The optimal $\sigma$ for
this run was $\sigma^\star = 0.398$ and the condition number at this
$\sigma^\star$ was $7\cdot 10^6$. The level sets of the reconstructed
potential are shown in Fig.~\ref{fig:2} and compared to those of the
original Mueller potential, while Fig.~\ref{fig:muelsweeptrj1}
compares the values of the original and reconstructed Mueller
potential along the TAMD trajectory shown in Fig.~\ref{fig:1}.  As a
simple estimate of the error, we used:
\begin{equation}
  \label{eq:error1}
    e_1 = \frac{\int_{\bar \Omega} 
    |V(\xx)-\tilde V(\xx)|dx}{\int_{\bar \Omega} 
    |V(\xx)|dx}
\end{equation}
where $\tilde V(\xx)$ denotes the reconstructed potential and
$\bar\Omega$ is the domain in which the original potential remains
less than $180$ above its minimum value. The error defined
in~(\ref{eq:error1}) for this calculation was $e_1=4.2\cdot10^{-3}$.

These results, which are already very good, can be improved by
diminishing $d$ and thereby increasing the number of centers without
having to increase the length of the TAMD trajectory.  For example, by
taking $d=0.12$, we obtained $351$ centers in a trajectory still $2
\cdot10^4$ steps long.  Using these centers to reconstruct the Mueller
potential, we obtained $e_1=3.2\cdot10^{-4}$. The level sets of the
reconstructed and original potential defined in Fig.~\ref{fig:2} now
overimposed so perfectly that they could not be distinguished on the scale
of Fig.~\ref{fig:2} (data not shown). The optimal $\sigma$ for this
calculation was $\sigma^\star=0.362$, at which value the residual was
$e_2(\sigma^\star)=4.7\cdot10^{-3}$ and the condition number was
$7\cdot10^{11}$.

Finally, we note that the result can also by improved by keeping the
same distance $d=0.12$ between centers but increasing the length of the
TAMD trajectory.  For instance, increasing the number of steps to
$5\cdot10^{4}$ produced a reconstruction with an error
$e_1=7.1\cdot10^{-5}$ (data not shown).

\section{Comparison with metadynamics}
\label{sec:metadynamics}

Because metadynamics~\cite{Laio02,Laio03} also uses an extended
dynamical system for $\xx$ and $\zz$ and the Gaussian
packet~(\ref{eq:kernel}) to represent $A(\zz)$, the single-sweep
method bears similarities with it.  Yet, there is an essential
difference between the two methods.  Unlike the single-sweep method,
metadynamics does not use the mean force, and estimates $A(\zz)$ by
direct sampling, which turns out to be a less efficient way to
proceed. Let us elaborate on this claim.

Recall that metadynamics uses an extended system like~(\ref{eq:tamd})
but where the equation for $\zz$ is replaced by~\cite{rem1}
\begin{equation}
  \label{eq:zzmeta}
  \begin{aligned}
    \gamma \dot \zz &= \kappa (\ttheta(\xx) - \zz) 
    + \sqrt{2 \gamma \beta^{-1}}\, \eeta(t) \\
    &\quad+ \nu\!\int_0^t\! \nabla_z \varphi_\sigma(|\zz(t) - \zz(t')|) dt'.
    \end{aligned}
\end{equation}
Here $1/\beta$ is now the physical temperature of the system, and
$\kappa$ and $\gamma$ are parameters playing the same role as
in~(\ref{eq:tamd}). The integral term in~(\ref{eq:zzmeta}) is a
flooding term (with $\nu>0$ controlling the flooding rate) which
deposits Gaussian packets~$\varphi_\sigma(u)$ on the energy 
landscape wherever $\zz(t)$ goes, thereby progressively leveling 
the effective free energy landscape felt by~$\zz(t)$. The negative of the integral of the
Gaussian packets deposited then gives an
approximation of the free energy.  For a trajectory with
$N_\text{max}$ time-steps of size~$\mathit{\Delta} t$, the
time-discretized approximation of this integral reads
(compare~(\ref{eq:radialbasis}))
\begin{equation}
  \label{eq:freemeta}
  \tilde A(\zz) = \nu\mathit{\Delta} t \sum_{n=1}^{N_\text{max}} 
  \varphi_\sigma(|\zz - \zz(n\mathit{\Delta} t)|) +C 
\end{equation}
where $C$ is a constant used to adjust the height of $\tilde
A(\zz)$.

\begin{figure}[t]
  \centerline{\includegraphics[width=3.25in]{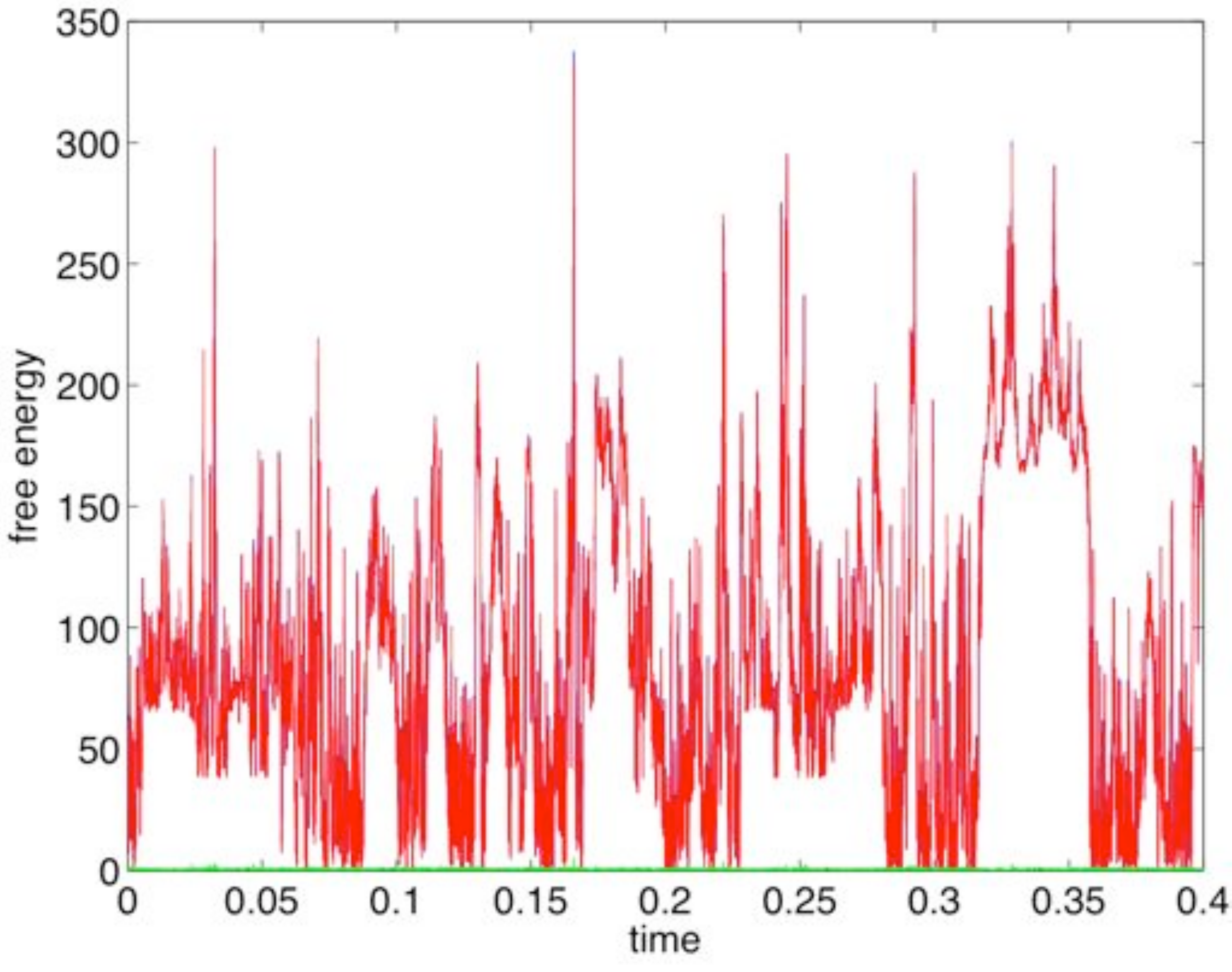}}
  \caption{% 
    Comparison between the
    original (blue line) and reconstructed Mueller potential (red
    line) computed along the $2\cdot 10^4$ steps single-sweep
    trajectory shown in Fig.~\ref{fig:1}. The green
    line shows the absolute value of the difference between the two.}
      \label{fig:muelsweeptrj1}
\end{figure}

Despite the fact that~(\ref{eq:freemeta}) uses Gaussian packets which
are radial-basis functions, the representation~(\ref{eq:freemeta}) is
very different from the standard radial-basis
representation~(\ref{eq:radialbasis}) used in the single-sweep method.
In particular, there are no coefficients $a_k$ to adjust
in~(\ref{eq:freemeta}). This has the important consequence that,
instead of requiring a single sweep across $\zz$-space to get an
accurate estimate of the free energy, metadynamics requires that the
trajectory revisits many times the same locations in $\zz$-space to
deposit centers (i.e. $N_\text{max}$ in~(\ref{eq:freemeta}) must be
much larger than $K$ in~(\ref{eq:radialbasis}) to achieve the same
accuracy). This is because the leveling out achieved by the integral
term in~(\ref{eq:zzmeta}) and, hence, the convergence of the
representation~(\ref{eq:freemeta}), only occur
statistically~\cite{lelievre,meta3} (in contrast, the mean force data
used at each center in the single sweep method contains already all
the statistical information needed at that center). This is consistent
with metadynamics being in essence an histogram method, albeit one
where the histogram windows are adjusted on-the-fly.

\begin{figure}[tbp]
  \centerline{\includegraphics[width=3.25in]{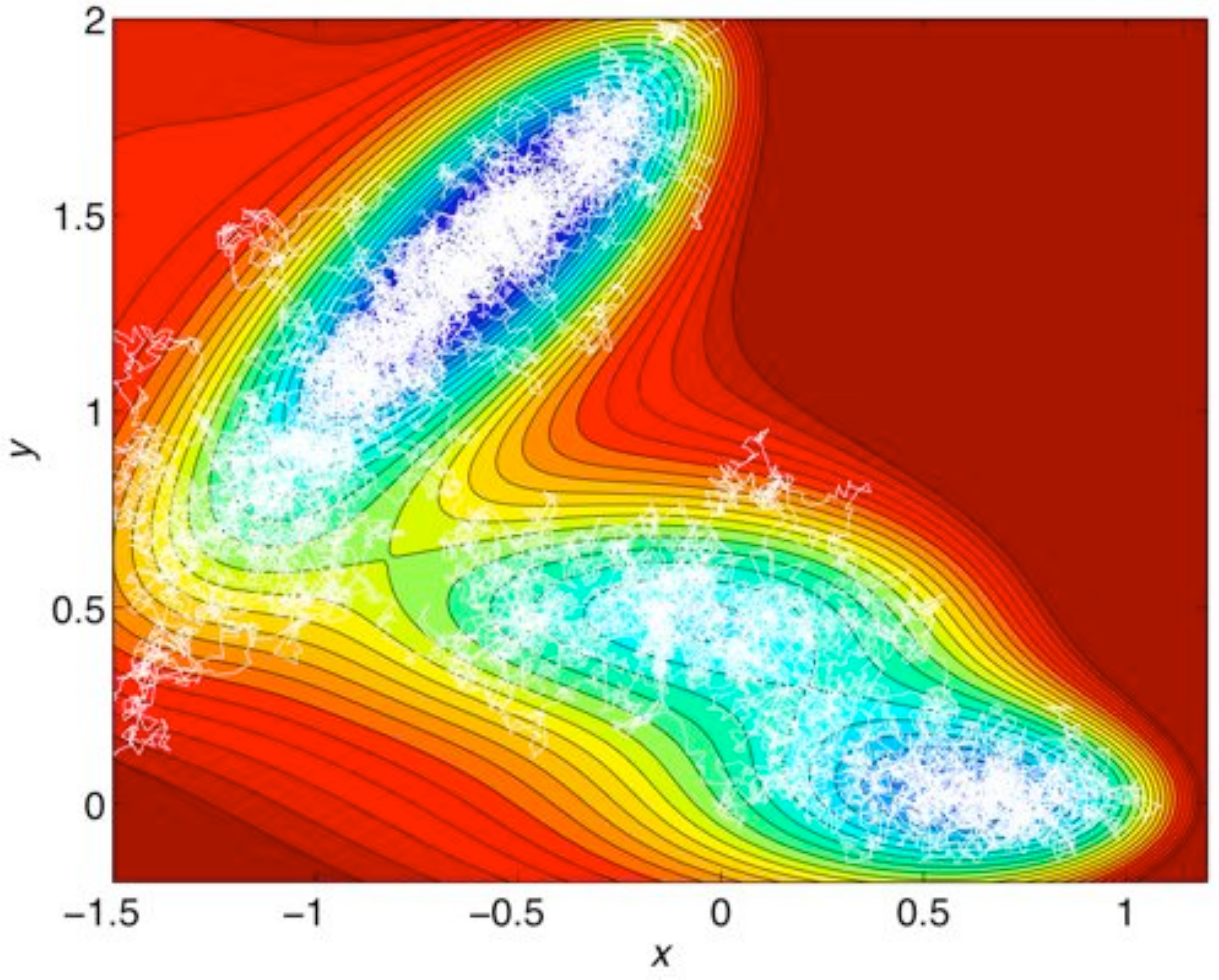}} 
  \caption{% 
    Metadynamics trajectory (white line) with $2\cdot10^4$ steps
    overimposed on the original Mueller potential.}
  \label{fig:muelmetaxy}
\end{figure}

\begin{figure}[t]
  \centerline{\includegraphics[width=3.25in]{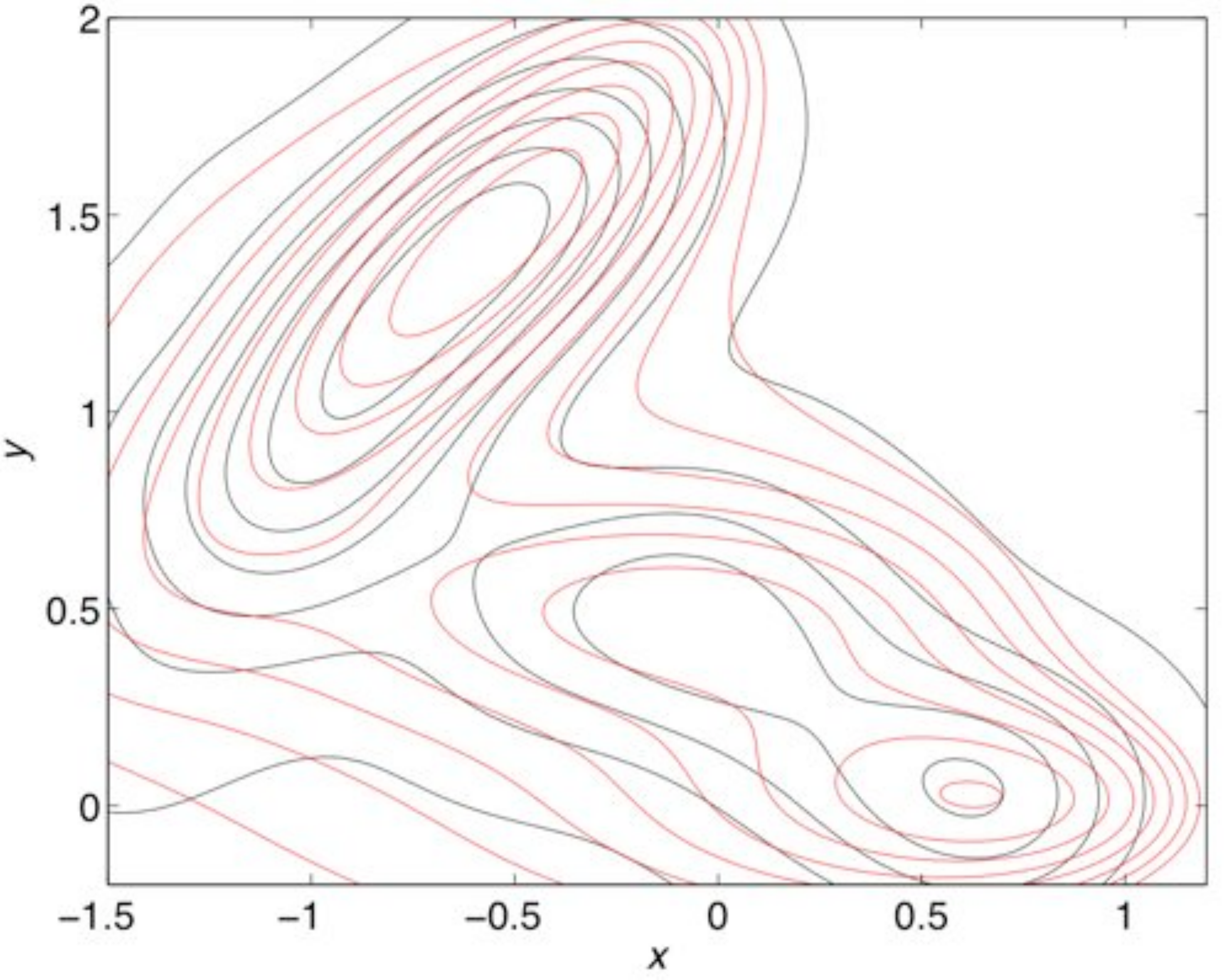}}
  \caption{%   
    Comparison between the level sets of the original Mueller
    potential (red curves) and the reconstructed potential using
    metadynamics with a trajectory of $2\cdot10^4$ steps with a time-step of
    ${\mathit\Delta }t = 2\cdot10^{-5}$ (same as in Figs.\ref{fig:1}
    and~\ref{fig:2}).  We only use 10 level sets evenly distributed
    between $V=0$ and $V=180$ because the differences between the maps
    are much bigger than with the single-sweep method and drawing more
    level sets makes the figure difficult to read. (There are only seven
    black level sets in the energy reconstructed by metadynamics
    because it levels off around $V=140$.)}
  \label{fig:6}
\end{figure}

\begin{figure}[t]
  \centerline{\includegraphics[width=3.25in]{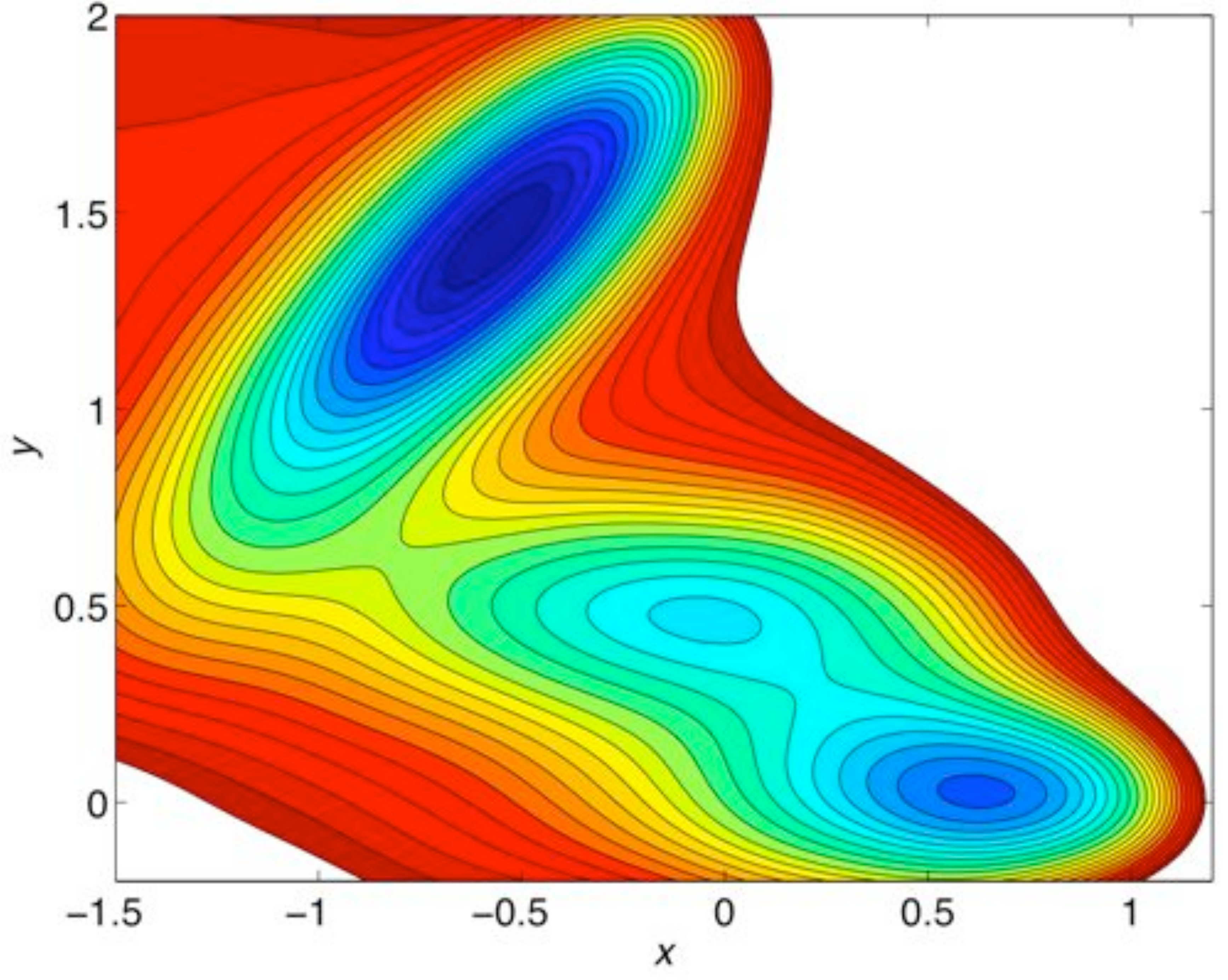}}
  \centerline{\includegraphics[width=3.25in]{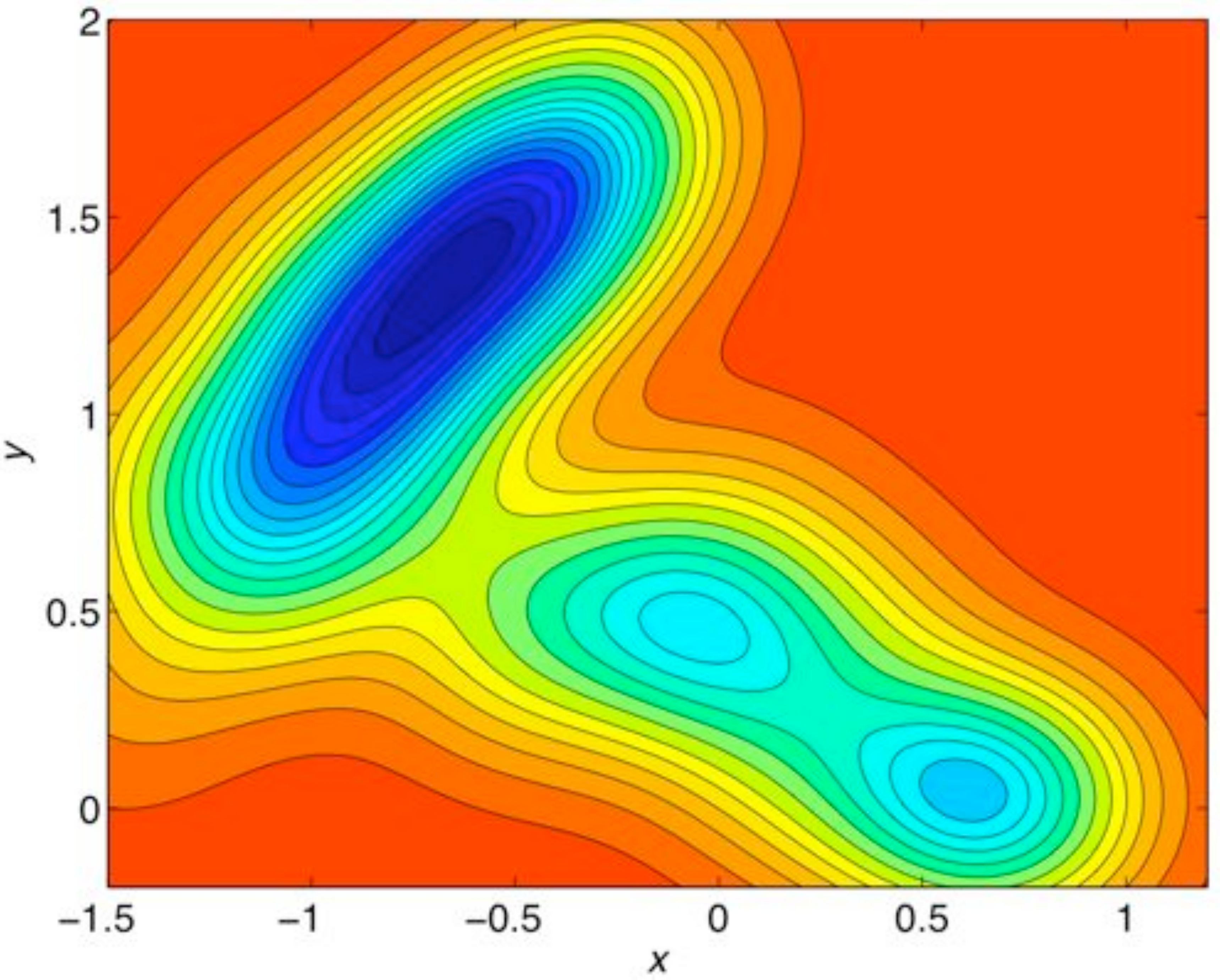}}
  \caption{% 
    Comparison between the Mueller potential as reconstructed by the
    single-sweep method (upper panel) and metadynamics (lower panel)
    both with a trajectory of $2\cdot10^4$ steps and a time-step of
    ${\mathit\Delta }t = 2\cdot10^{-5}$. Other representation of these
    contourplots were already shown in Figs.~\ref{fig:2}
    and~\ref{fig:6} respectively. The map reconstructed by the
    single-sweep method is very close to the map of the original
    Mueller potential.  The colormaps used in both panels are the same
    and the reconstructed potentials are shifted so that their minimum
    is $V=0$; the white region in the left panel is where the energy
    is above $180$ and is not shown (the result of metadynamics shown
    in the right panel levels off around $V=140$ which is why there is
    no white).}
  \label{fig:6b}
\end{figure}

\begin{figure}[tbp]
  \centerline{\includegraphics[width=3.25in]{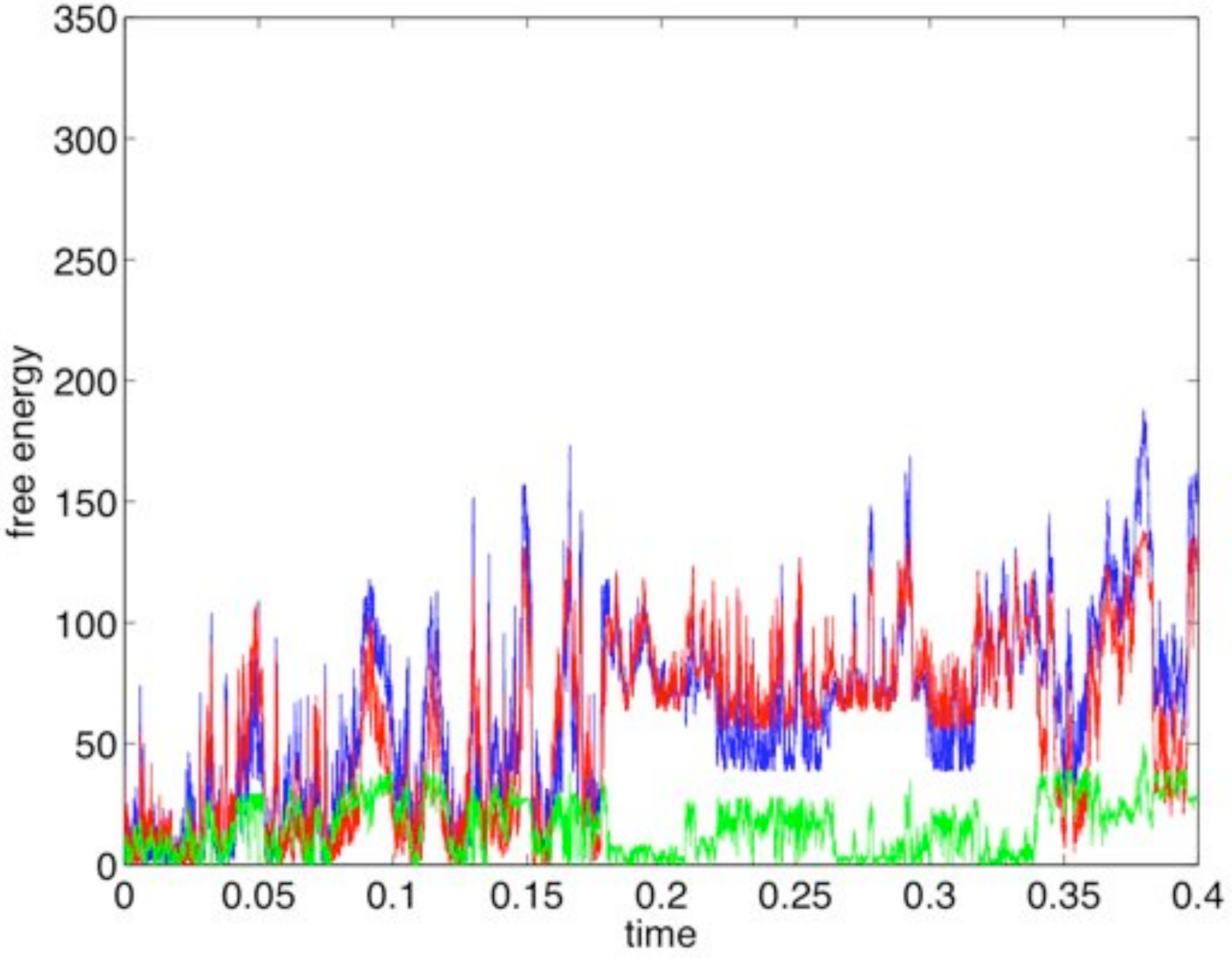}} 
  \caption{% 
    Comparison of the original (blue line) and reconstructed Mueller
    potential (red line) computed along the $2\cdot10^4$ steps
    metadynamics shown in Fig.~\ref{fig:muelmetaxy}. The green line
    shows the absolute value of the difference between the two.  The
    green line here should be compared with the one in
    Fig.~\ref{fig:muelsweeptrj1} for the single-sweep method: the
    discrepancy between the original and reconstructed potential is
    always larger with metadynamics than with the single-sweep
    method.}
      \label{fig:muelmetatrj1}
\end{figure}

What this entails in terms of efficiency can be illustrated on the
two-dimensional Mueller example considered before.  In this example,
to generate the metadynamics trajectory we used~(\ref{eq:tamdex}) with
flooding terms added as in~(\ref{eq:zzmeta}), consistent with what was
done in Ref.~\cite{meta3} to test the efficiency of metadynamics in a
similar set-up.  Fig.~\ref{fig:muelmetaxy} shows the metadynamics
trajectory obtained by integrating~(\ref{eq:zzmeta}) for
$2\cdot10^4$ timesteps with a time-step of $\mathit{\Delta }
t=2\cdot10^{-5}$ (same as in the single-sweep method for the results
shown in Figs.~\ref{fig:1} to~\ref{fig:muelsweepres}).  As can be seen
in Fig.~\ref{fig:muelmetaxy}, this number of timesteps was enough for
the trajectory to visit the important regions of the potential. The
reconstructed free energy from this calculation is compared in
Fig.~\ref{fig:6} to the original one.  The metadynamics result is also
compared in Fig.~\ref{fig:6b} to the one obtained by the single-sweep
method with $174$ centers.  The error~(\ref{eq:error1}) for this
metadynamics calculation was $e_1=0.16$, i.e. almost two orders of
magnitude higher than with the single-sweep method.
Fig.~\ref{fig:muelmetatrj1} shows the original and reconstructed
Mueller potential along the metadynamics trajectory (blue and red
lines, respectively), together with the absolute value of their
difference (green line).  By comparing Figs.~\ref{fig:muelsweeptrj1}
and~\ref{fig:muelmetatrj1}, it can be seen that the discrepancy
between the original and reconstructed potential is much larger with
metadynamics than with the single-sweep method on trajectories of the
same length. Note that these results clearly indicates that it is not
sufficient that the metadynamics trajectory visits once a region on
phase space to get an accurate representation of the free energy in
this region. This was already noted in Refs.~\cite{lelievre,meta3}.

We were able to improve the metadynamics result by extending the
simulation to $2\cdot10^{5}$ steps. The covering of the important
regions in the potential was now extensive (data not shown), and the
reconstructed potential (data not shown) looked visually better than
the one obtained with the shorter trajectory. Yet the
error~(\ref{eq:error1}) was $e_1=0.12$, i.e. still two orders of
magnitude larger than the highest error we obtained with the
single-sweep method using a 10 times shorter trajectory. We did not
attempt to go to longer runs with metadynamics because the memory term
in~(\ref{eq:zzmeta}) makes such simulations increasingly expensive
(their cost scales as the square of the number of timesteps).  It
should also be stressed that in all these calculations, we optimized
the parameters $\nu$, $\sigma$ and $1/\beta$ the best we could. This
optimization, however, turns out to be complicated since there is no
systematic way to perform it because, unlike with the single-sweep
method, there is no objective function to minimize in metadynamics.
The results shown in
Figs.~\ref{fig:muelmetaxy}--\ref{fig:muelmetatrj1} were obtained with
$\nu = 2\cdot10^3$, $\sigma=0.2$ and $1/\beta=10$.

To be fair, we should conclude this comparison by mentioning that the
simplicity of the Mueller potential example tends to exaggerate the
gain that the single sweep method provides over metadynamics. Indeed,
in realistic situations, the single-sweep method also requires to
compute the mean force via~(\ref{eq:meanforceapprox}), an operation
which was unnecessary in the Mueller example since the force was
readily available. Computing the mean force adds an extra cost to the
method.  It is worth stressing again, however, that the computation of
the time averages in~(\ref{eq:meanforceapprox}) can be distributed
over several processors. This means that in the ideal situation where
the user has at least one processor per center, the effective time to
compute all of the mean forces is the same as the one for computing a
single one of these forces, i.e. we have perfect scalability.
Metadynamics can be parallelized per replica as well, as was proposed
in Refs.~\cite{metap1,metap2,metap3}, but not as straightforwardly and
not with perfect scalability. Indeed, in all of these versions of
metadynamics, the simulated replica are never completely independent
from each other.

\section{Free energy of alanine dipeptide in solution}
\label{sec:AD}

In this section, we use the single-sweep method to reconstruct the
free energy of the solvated alanine dipeptide (AD) molecule in two and
four torsion angles at $300$~K.  While AD is not an example of
biochemical interest \textit{per~se}, we study it because it has been
extensively used as a benchmark example for free energy calculations
in the literature~\cite{karplus, tuckad, meta_ad}. On top of this the system 
is simple enough that we can use it to systematically investigate how the
accuracy of the reconstruction method depends on the number of centers
and how robust the method is with respect to statistical errors in the
input data for the mean forces.  Another question we investigate in
this section is the robustness of the method against the choice of
radial-basis functions.  Specifically, we compare results obtained
using the Gaussian packet~(\ref{eq:kernel}) and the Wendland function
\begin{equation}
  \label{eq:wend}
  \varphi(u) = {(1-u)^6_+(35u^2+18u+3)}
\end{equation}
where $(f(u))_+=f(u)$ if $f(u)>0$, and $(f(u))_+=0$ otherwise.
(\ref{eq:wend}) is another well-known example of radial-basis function
which has the pleasant property that it is compactly supported. This
property is appealing in the calculations since it limits the range
over which centers interact in~(\ref{eq:objective}).

All MD simulations reported below were performed with a
version of the MOIL code~\cite{moil} suitably modified by us, 
and the AMBER/OPLS~\cite{ff} force field (for
details of the MD set-up see Appendix~\ref{sec:MD}).

\subsection{Two angles calculation} 
\label{sec:2angles}

\begin{figure}[t]
  \centerline{\includegraphics[width=3.25in]{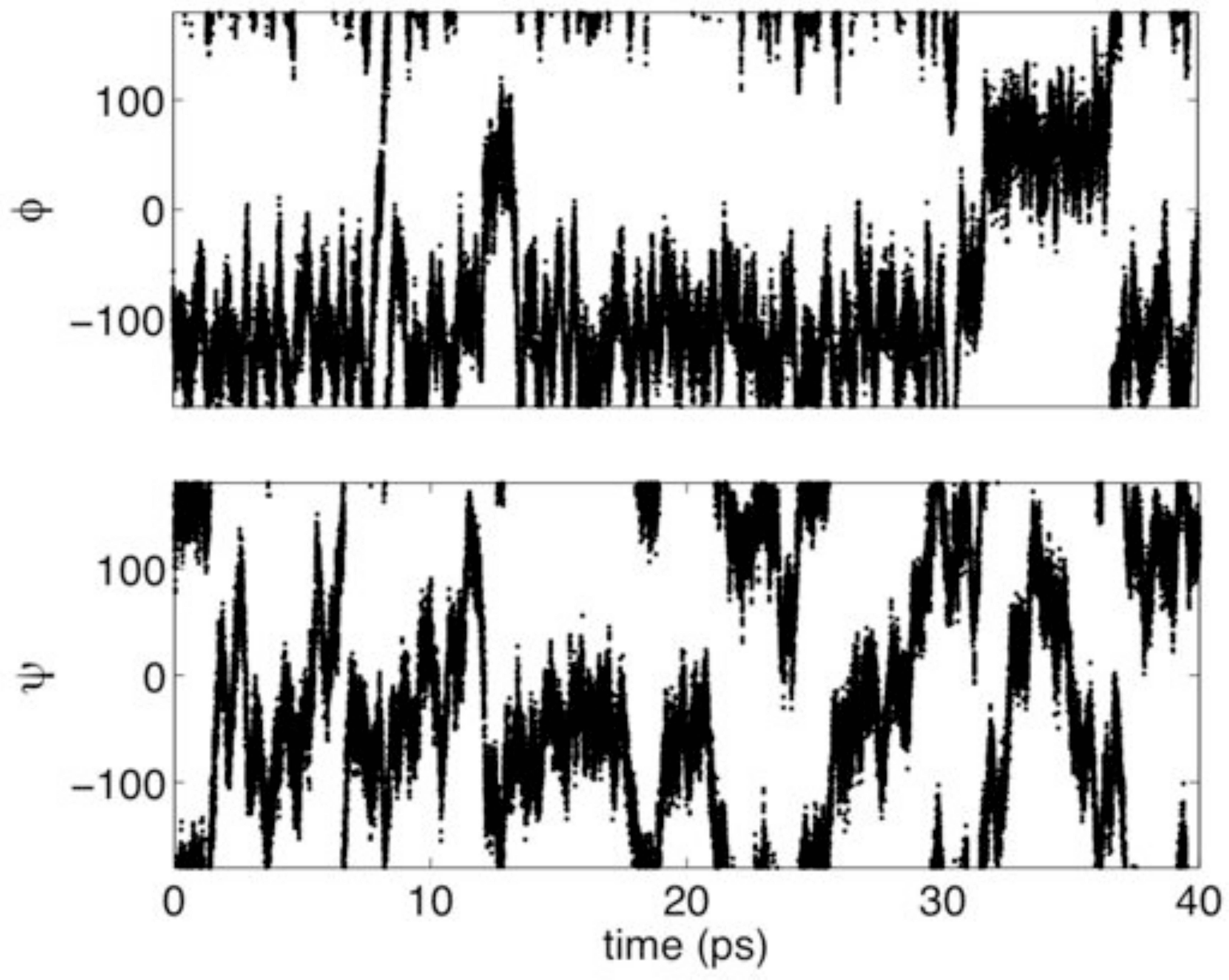}}
  \caption{% 
    Time series of the dihedral angles $\phi$ and $\psi$ along the
    $40$~ps long TAMD trajectory for the solvated alanine dipeptide
    (AD).  }
      \label{fig:adtrj}
\end{figure}

We use the standard dihedral angles $\phi$ and $\psi$.  At~$300$~K,
the system is confined in a region of the $(\phi,\psi)$ space with
$\phi < -50^\circ$ by energy barriers higher than $1/\beta$.  In order
to overcome these barriers and sweep through the whole
$[-180^\circ,180^\circ]^2$ space, we generated a trajectory by
using~(\ref{eq:tamd}) with $(z_1,z_2)=(\phi,\psi)$,
$\kappa=100$~kcal/mol/rad$^2$, a friction coefficient
$\gamma=0.5$~kcal$\times$ps/mol/rad$^2$ and an artificial temperature
$1/\bar\beta =9.5$~kcal/mol.  With this choice of the parameters, the
important regions of the $[-180^\circ,180^\circ]^2$ space were visited
in $4\cdot10^4$ steps ($40$~ps in the time units of the MD variables).
The time series of $\phi$ and $\psi$ along this trajectory are shown in
Fig.~\ref{fig:adtrj}. Variations in $\kappa$, $\gamma$ and $1/\bar
\beta$ led to qualitatively similar TAMD trajectories, indicating that
the the method is robust with regard to the choice of these
parameters. Sets with a different number~$K$ of centers were deposited
along the TAMD trajectory afterwards by processing this trajectory
using various distances~$d$ between the centers.  Specifically, we
generated sets of $90$, $128$, $151$, $188$, $219$ and $262$ centers
using, respectively, $d=31.77^\circ$, $26.00^\circ$, $23.87^\circ$,
$21.37^\circ$, $20.00^\circ$, $17.92^\circ$.

Given a set of $K$ centers $(\phi_k,\psi_k)$, we computed the mean
forces via $K$ independent MD simulations with restraints at
$(\phi,\psi)=(\phi_k,\psi_k)$, i.e. by simulating~(\ref{eq:tamd2}) in
the isokinetic ensemble at $300$~K, and estimated the mean force
via~(\ref{eq:meanforceapprox}) with $\bar\kappa=100$~kcal/mol/rad$^2$.
This value of $\bar \kappa$ was high enough since we checked that the
reconstructed free energy remained invariant with higher values
of~$\bar \kappa$ (we did so up to $\bar\kappa=10^3$~kcal/mol/rad$^2$).
We then used this data in the reconstruction procedure explained in
Sec.~\ref{sec:reconstruct}. Note that since the free energy in the
$(\phi,\psi)$ angle is periodic, we have to periodically extend the
centers for the representation. This amounts to changing the
representation in~(\ref{eq:radialbasis}) into
\begin{equation}
  \label{period2}
  \tilde A(\zz) = \sum_{\hat{\uu{n}}\in Z^N}\sum_{k=1}^K 
  a_k \varphi_\sigma(|\zz- 
  \zz_k+2\pi~\hat{\uu{n}}\cdot \hat{\uu{e}}|) +C,
\end{equation}
where $\hat{\uu{e}}$ is the unit vector in $\RR^N$.  In practice, only
few periodic replica of the centers are needed (i.e. $\hat n_i=2$ for
$i=1,\dots,N$) because the radial-basis functions centered at the
centers further away from the cell under consideration make negligible
contributions to the result in this cell.

\subsubsection{Calculation with $d=21.37^\circ$ ($188$ centers) 
  and $T=50$~ps.}  
\begin{figure}[t]
  \centerline{\includegraphics[width=3.25in]{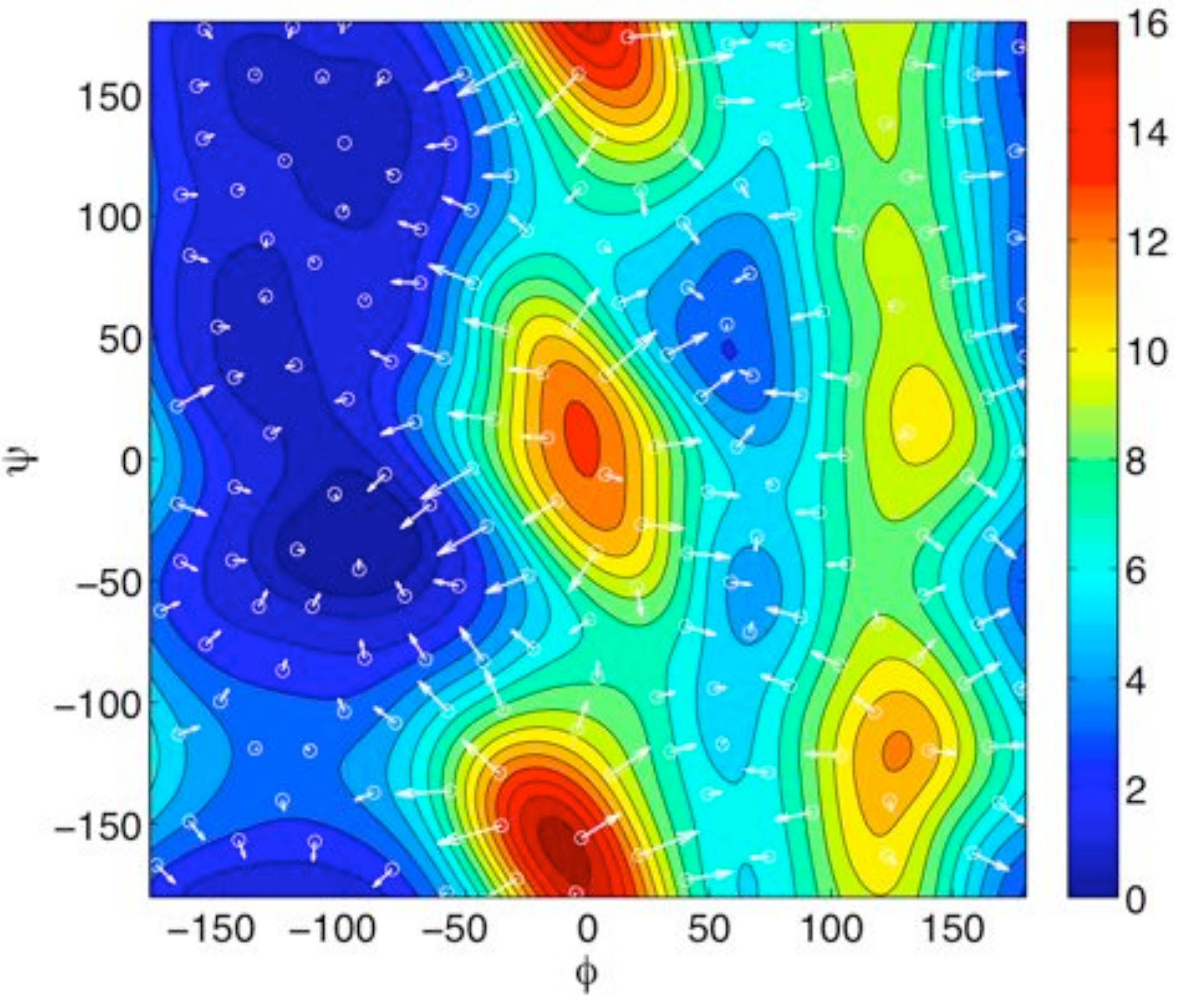}}
  \caption{%   
    Free energy of AD in the $\phi$ and $\psi$ dihedral angles at
    $300$~K calculated with the single-sweep method by using $188$
    centers deposited at a distance of $d=21.37^\circ$ from each
    other. Units for the free energy are kcal/mol, and contour levels
    are plotted at $0.5$~kcal/mol, $1$~kcal/mol, and then every
    $1$~kcal/mol.  The optimal $\sigma$ in this reconstruction was
    $\sigma=44.73^\circ$.  The centers are represented as white
    circles. At every center, the corresponding mean force vector is
    also shown.  Mean forces were calculated by
    using~(\ref{eq:meanforceapprox}) with
    $\bar\kappa=100$~kcal/mol/rad$^2$ and $T=50$~ps.}
  \label{fig:7}
\end{figure}

We first detail our result with this choice of
parameters to pick an example which led to a good balance between
accuracy and efficiency.  Other choices of parameters are discussed
below. Thus, Fig.~\ref{fig:7} shows the reconstructed free energy map
obtained with $d=21.37^\circ$ ($188$ centers) and by computing the
mean forces from~(\ref{eq:meanforceapprox}) with $T=50$~ps. The
optimal $\sigma$ in this calculation was $\sigma^\star=44.73^\circ$.
In the figure, the minimum of the free energy is set at $0$~kcal/mol,
and contour levels are plotted at $0.5$~kcal/mol, $1$~kcal/mol and
then every $1$~kcal/mol.  The centers are represented as white
circles, and the mean forces at the centers as arrows.

\begin{figure}[t]
  \centerline{\includegraphics[width=3.395in]{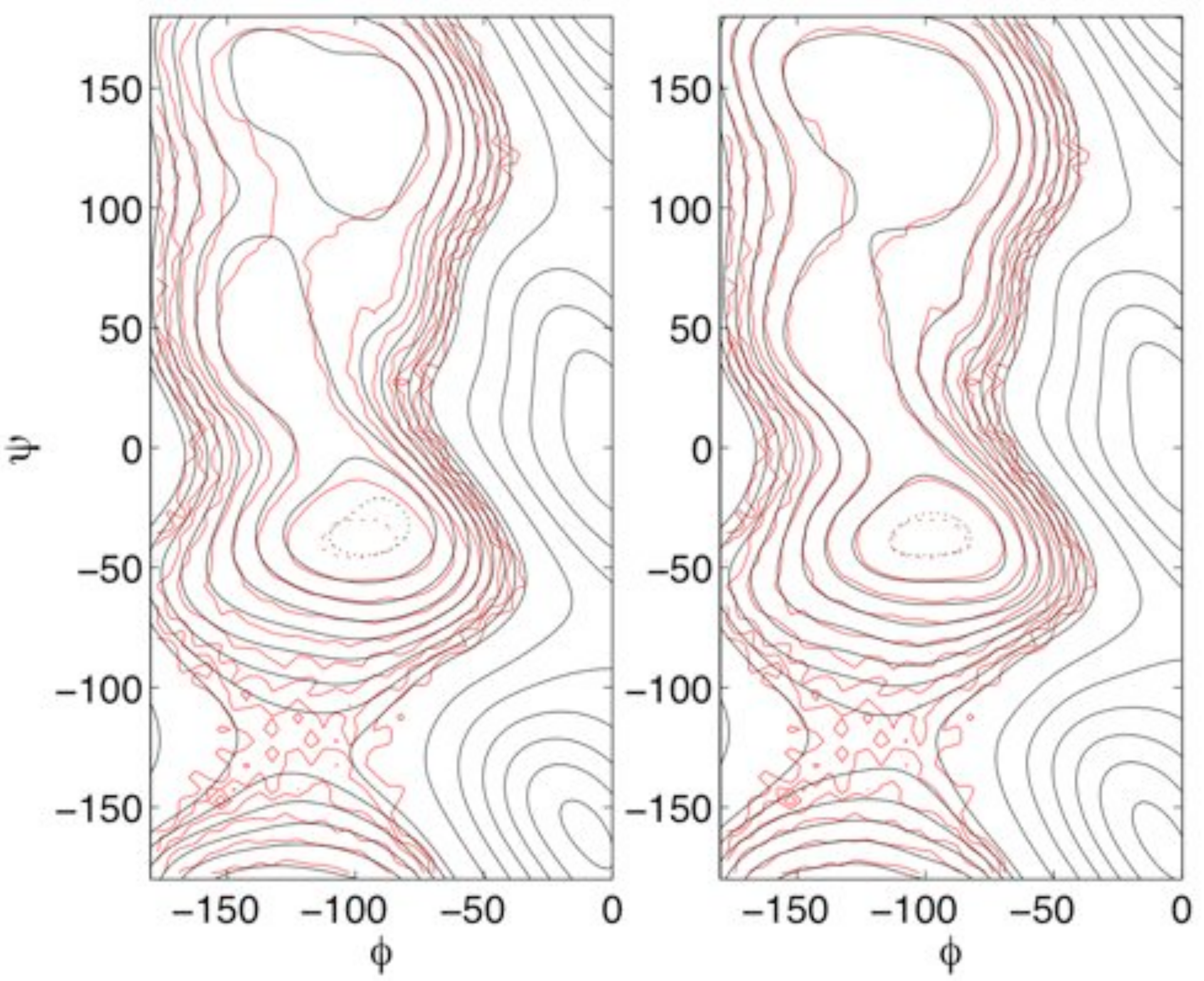}}
  \caption{%  
   Comparison between the free energy obtained by single-sweep (black
    lines) method and DMDS (red lines) for AD. The left panel shows
    the result with with $188$ centers, the right panel the one with
    $262$ centers.  The contour levels of the free energy are plotted
    at $0.1$~kcal/mol (dotted lines), from $0.5$ to $4.0$~kcal/mol
    separated by $0.5$~kcal/mol, and then separated by $2$~kcal/mol.}
  \label{fig:8}
\end{figure}

Since the free energy map depends on the force field used, comparison
with results in the literature is difficult. To assess the accuracy of
our result self-consistently, we compared it with the free energy
calculated by computing the PDF of $\phi$ and $\psi$ from a direct MD
simulation (DMDS) of about $30$~ns.  While this trajectory does not
cover all the $[-180^\circ,180^\circ]^2$ space, it covers the
important regions and allows for an unbiased estimation of the free
energy in these regions which can be used as benchmark. The left panel
in Fig.~\ref{fig:8} shows the contour levels of the free energy from
the single-sweep (black lines) and that from DMDS (red lines). The
contour levels are plotted at $0.1$~kcal/mol (dotted lines), every
$0.5$~kcal/mol from $0.5$ to $4$~kcal/mol, and then every
$2$~kcal/mol. As can be seen, single-sweep results agree remarkably
well with those of the DMDS.

In terms of cost, to generate the result shown in Fig.~\ref{fig:7}, we
had to make one simulation run of $40$~ps to generate the TAMD
trajectory, plus $188$ independent runs of $50$~ps distributed on
different nodes (the additional cost of estimating the parameters
$a^\star_k$ and $\sigma^\star$ to use in~(\ref{eq:radialbasis}) is
insignificant). This makes for a total of $9.4$~ns of absolute
simulation time.  However, after distribution, the effective
simulation time needed is only $90$~ps.  On top of this, we show below
that a good estimate of the free energy can be obtained with as low as
90 centers (i.e. with an absolute simulation time of $4.5$~ns and the
same effective simulation time,~$90$~ps).  For comparison, in
Ref.~\cite{meta_ad} Ensing~\textit{et al.} report a $4$~ns calculation
performed with metadynamics to estimate the free energy of AD in
$\phi$ and $\psi$. It is not clear how well this metadynamics
calculation can be parallelized to reduce its effective cost (it was
not parallelized in Ref.~\cite{meta_ad}). In addition, the result of
this metadynamics calculation is unlikely to be as accurate as the one
in Fig.~\ref{fig:7} (in Ref.~\cite{meta_ad} no comparison like the one
shown in Fig.~\ref{fig:8} is provided)

\subsubsection{Robustness and convergence analysis}

Next we analyze how robust are the results with respect to the
statistical error in the mean force data and the choice of
radial-basis function. We also analyze convergence in function of the
number of centers. As reference value, we take the free energy
reconstructed with $d=17.92^\circ$ ($262$ centers) and $T=250$~ps of
time averaging in~(\ref{eq:meanforceapprox}).  The map of the free
energy calculated with these parameters (data not shown) is visually
very similar to the one shown in Fig.~\ref{fig:7}, but it is more
accurate.  The residual error can be estimated from the right panel of
Fig.~\ref{fig:8} which shows the contour levels of the free energy
from the single-sweep (black lines) and that from DMDS (red lines):
these level sets coincide up to statistical errors in the DMDS,
indicating that the free energy provided by the single sweep method
with $262$ centers and $T=250$~ps can indeed be taken as an ``exact''
benchmark.

\begin{figure}[tbp]
 \centerline{\includegraphics[width=3.25in]{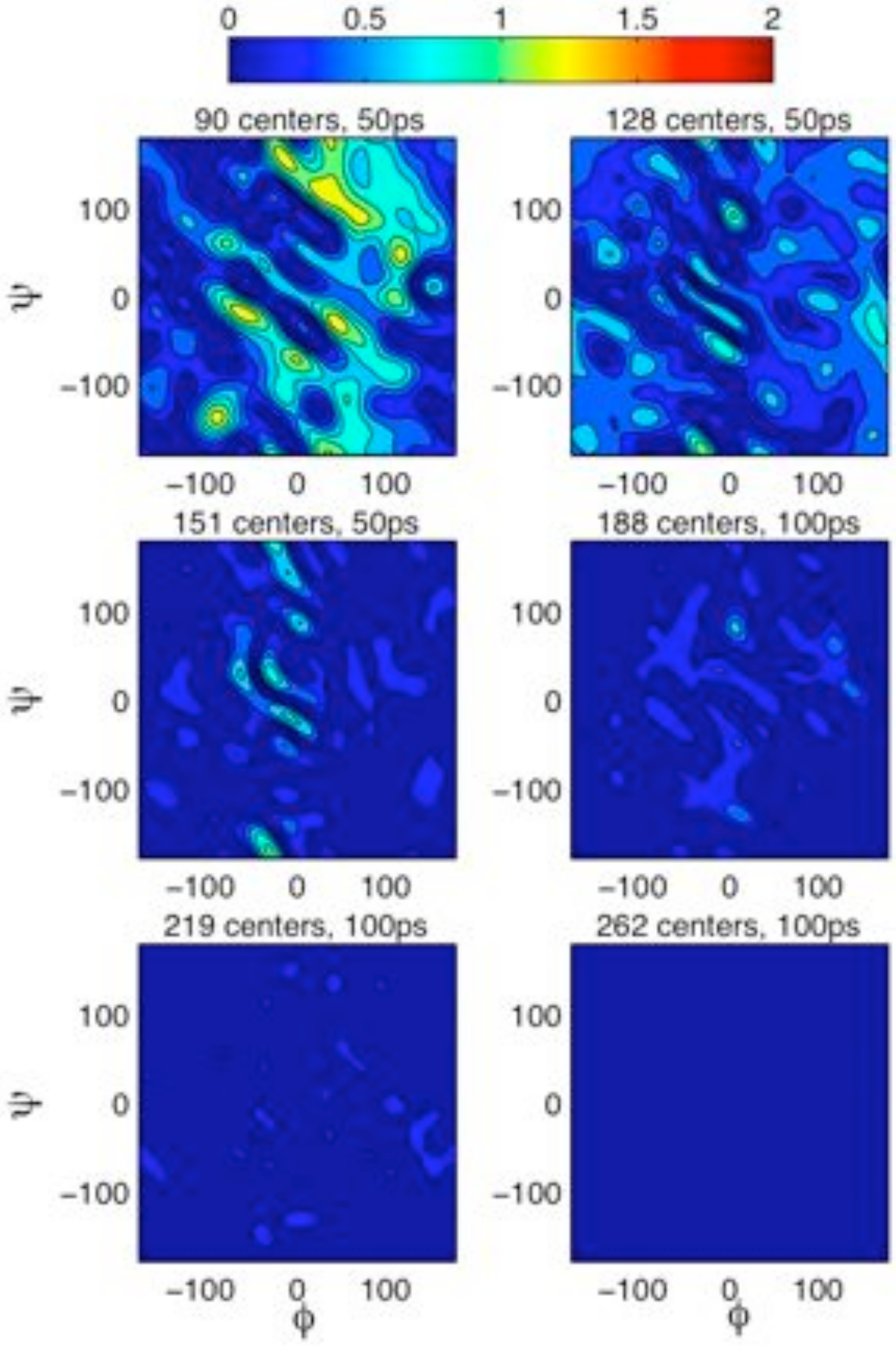}}
  \caption{%   
    Difference maps with respect to the AD free energy in $\phi$ and
    $\psi$ reconstructed with Gaussian functions using $262$ centers
    and $T=250$~ps. The figures in the different panels correspond to
    various number of centers and length of time averaging for the
    mean forces, as indicated.  Units are kcal/mol.  Note that the
    scale of the colormap is different from the one in
    Fig.~\ref{fig:7}. In particular, the differences are mostly
    below $0.5$~kcal/mol with $151$ centers and $50$~ps simulations
    already.}
      \label{fig:diffsad}
\end{figure}

Fig.~\ref{fig:diffsad} shows the differences between the map of the
reference free energy reconstructed with $d=17.92^\circ$ ($262$
centers) and $T=250$~ps and those reconstructed with less centers and
shorter restrained simulations. The largest errors are in the regions
corresponding to the highest peaks of the free energy (these are also
the regions were the least centers were deposited).  The differences
never exceed $1.25$~kcal/mol with $90$ centers and $T=50$~ps and they
fall mostly below $0.5$~kcal/mol with $151$ centers and $T=50$~ps already.

\begin{figure}[t]
 \centerline{\includegraphics[width=3.25in]{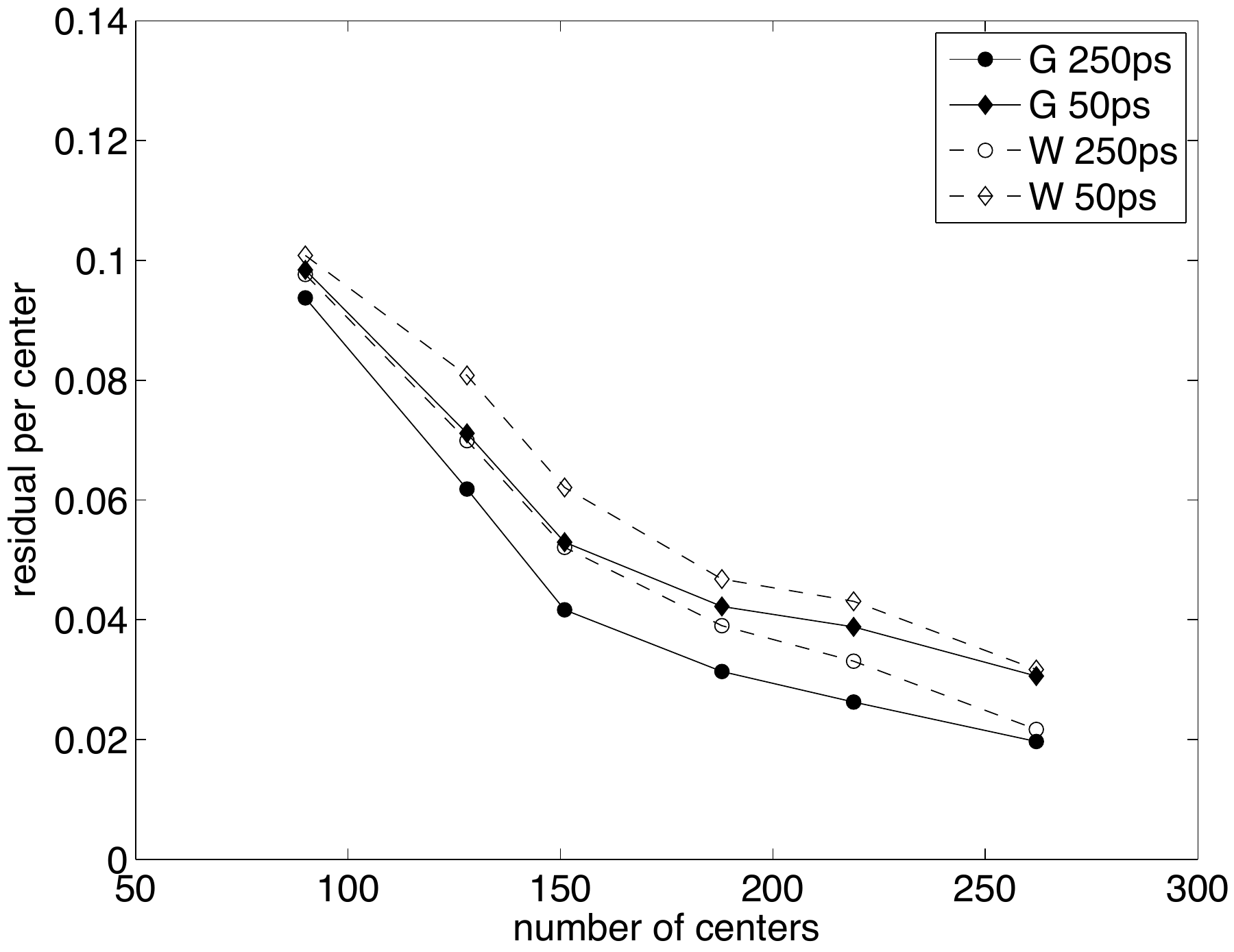}}
  \caption{% 
    Residual per center versus number of centers in the reconstruction
    of the AD free energy in $\phi$ and $\psi$ angles.  Data are from
    calculations with different time averaging length, using Gaussian
    (G) and Wendland (W) basis functions.}
      \label{fig:resad}
\end{figure}

We also compared the quality of the reconstruction of the free energy
when using Gaussian~(\ref{eq:kernel}) and Wendland~(\ref{eq:wend})
basis functions. Fig.~\ref{fig:resad} shows the residual per center
versus the number of centers for AD, by using Gaussian (filled
symbols) and Wendland (empty symbols) basis functions, using $T=50$~ps
(diamonds) and $T=250$~ps (circles) long restrained simulations to
estimate the mean forces. At equal values of $K$ and $T$, the
reconstruction is slightly more accurate with Gaussian than with
Wendland functions, though these differences turn out to be quite
small in terms of the free energy maps themseves (data not shown).

Using longer simulations for the mean force (which means a smaller
random error on these forces) also improves the results. With $262$
centers and $T=250$~ps long simulations for the mean forces, the maps
reconstructed with Gaussian and Wendland basis functions were almost
identical (data not shown), and they were not significantly different
from the map shown in Fig.~\ref{fig:7}. These results, however, were
obtained at very different values of optimal~$\sigma$:
$\sigma^\star=26.33^\circ$ with Gaussians and
$\sigma^\star=123.20^\circ$ with Wendland functions. The condition
numbers at the optimal $\sigma$ were $3889.67$ and $868.86$,
respectively, which is low. Note that the same trends here described
were observed in a periodic test case for which the potential was 
known exactly (data not shown).

\subsection{Four angles calculation}
\label{sec:4angles}

As a second more challenging test, we computed the free energy of AD
in the four torsion angles, $\phi$, $\psi$, $\theta$, and~$\zeta$.  A
TAMD trajectory of $44$~ps was generated by using~(\ref{eq:tamd}) with
$(z_1,z_2,z_3,z_4)=(\phi,\psi,\theta,\zeta)$,
$\kappa=100$~kcal/mol/rad$^2$, an artificial temperature $1/
\bar\beta=9.5$~kcal/mol, friction coefficients
$\gamma=0.5$~kcal$\times$ps/mol/rad$^2$ for $\phi$ and $\psi$ and
$\gamma=1$~kcal$\times$ps/mol/rad$^2$ for $\theta$ and $\zeta$. The MD
potential keeps the amide planes in \textit{trans} configuration, and
so $\theta$ and $\zeta$ were varying in the range
$[-70^\circ,70^\circ]$. In $44$~ps, the TAMD trajectory covered well
the accessible state space for $\phi$, $\psi$, $\theta$ and $\zeta$,
in the sense that the time series for these angles were similar in
their respective state space to those shown in Fig.~\ref{fig:adtrj}
(notice however that extensive coverage of the four-dimensional space
is unlikely in so short a run).  Along the TAMD trajectory, $200$
centers at a distance of $d=45.84^\circ$ were deposited.  At these
centers, the mean forces $\ff_k$ were computed by
using~(\ref{eq:meanforceapprox}) with
$\bar\kappa=100$~kcal/mol/rad$^2$ and $T=50$~ps (i.e. the absolute
time of simulation was about $10$~ns, but the effective time after
distribution was $94$~ps only). We used these $\ff_k$ in the objective
function~(\ref{eq:objective}) to finally get the
representation~(\ref{eq:radialbasis}) of the four-dimensional free
energy~$\tilde A(\phi,\psi,\theta,\zeta)$. The optimal $\sigma$ in
this representation was $\sigma^\star=67.63^\circ$ and the condition
number at this value of $\sigma$ was $4214.05$.

\begin{figure}[t]
  \centerline{\includegraphics[width=3.25in]{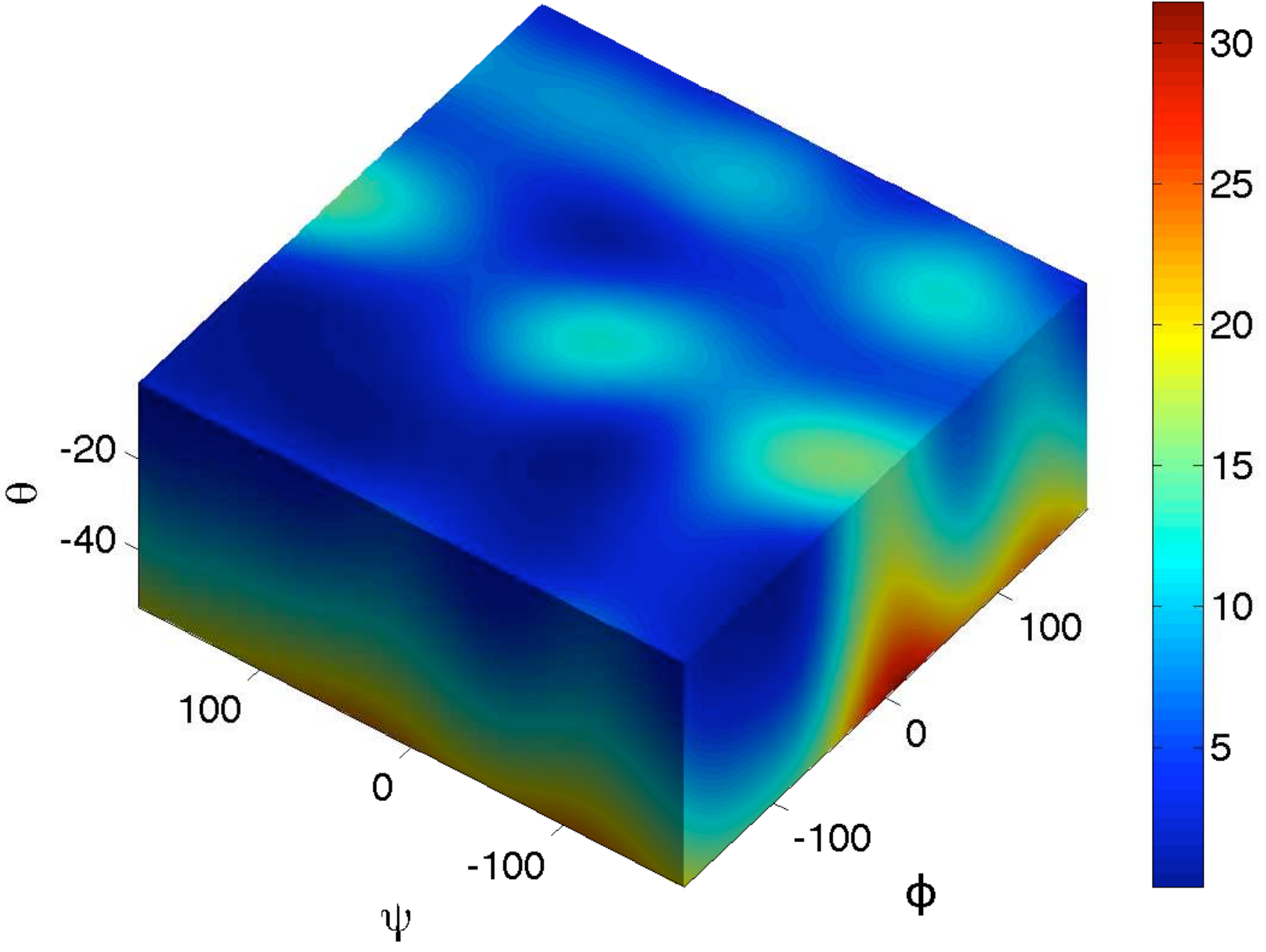}}
  \caption{%   
    Free energy of AD in the $\phi$, $\psi$, $\theta$ angles obtained
    from the marginal in these angles of the PDF associated with the
    free energy in four angles $\tilde A(\phi,\psi,\theta,\zeta)$.
    Data are represented for $\theta\in[-50^\circ,-3^\circ]$. 
    Note that the scale of the colormap is different from the one in
    Fig.~\ref{fig:7}}
  \label{fig:9}
\end{figure}

\begin{figure}[t]
  \centerline{\includegraphics[width=3.25in]{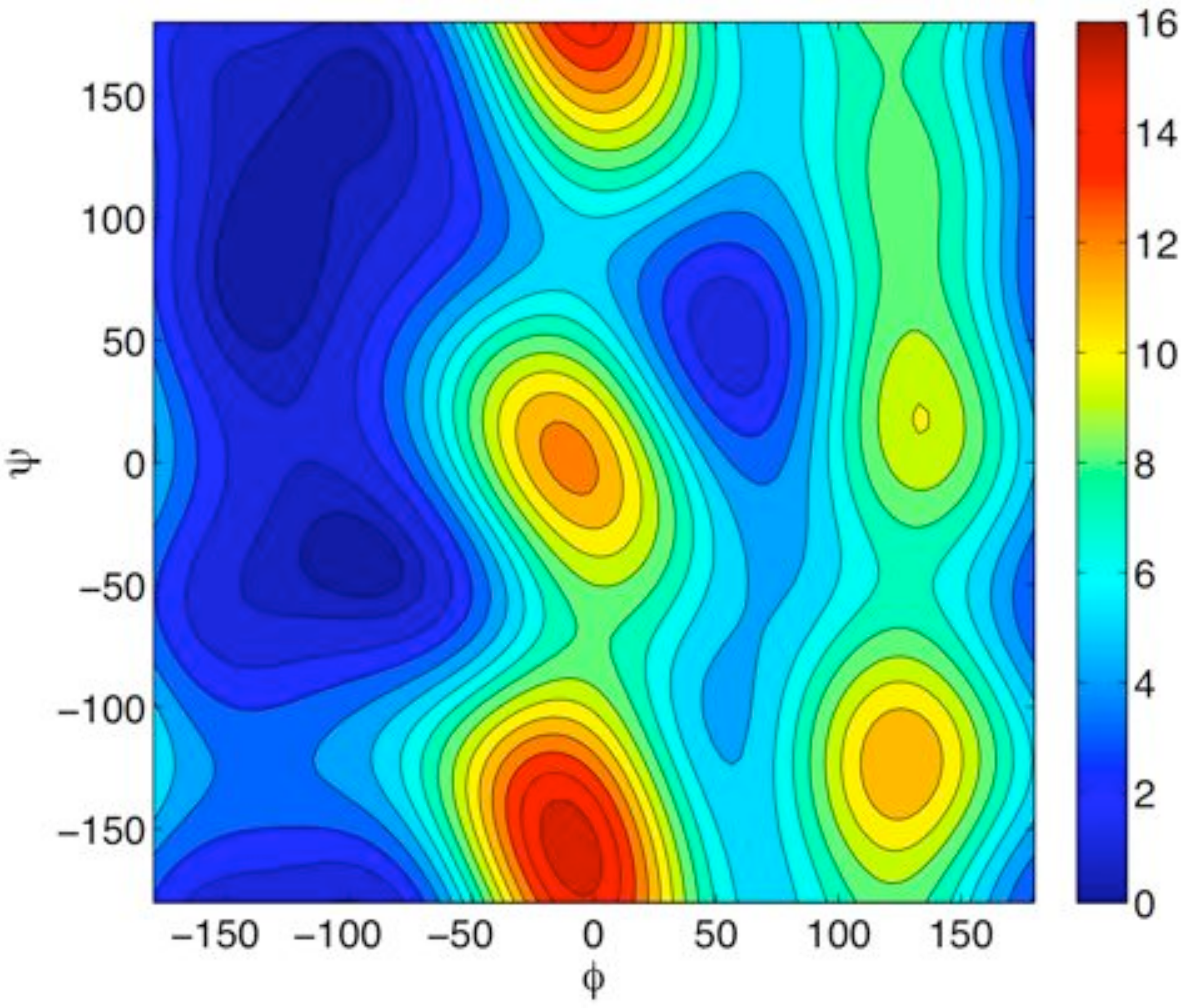}}
  \caption{%   
    Free energy of AD in the $\phi$, $\psi$ angles obtained from the
    marginal in these angles of the probability density associated
    with the free energy in four angles $\tilde A(\phi,\psi,\theta,\zeta)$.
    Contour levels are as in Fig.~\ref{fig:7}. Note the remarkable
    agreement between this map and the one shown in
    Fig.~\ref{fig:7}.}
  \label{fig:10}
\end{figure}

Since a full graphical representation of~$\tilde
A(\phi,\psi,\theta,\zeta)$ is not possible, we did several tests to
validate our result.  Fig.~\ref{fig:9} shows the three dimensional
free energy $\tilde A(\phi,\psi,\theta)$ obtained from the marginal in
these angles of the PDF associated with $\tilde
A(\phi,\psi,\theta,\zeta)$. This marginal was calculated
\textit{a~posteriori} by numerical integration over $\zeta$ of
$e^{-\beta \tilde A(\phi,\psi,\theta,\zeta)}$ with the full $\tilde
A(\phi,\psi,\theta,\zeta)$ reconstructed by the single-sweep method.
The map is reasonable, and shows nontrivial features in all three directions.
Fig.~\ref{fig:10} shows the two dimensional free energy $\tilde
A(\phi,\psi)$ obtained from the marginal in these angles of the PDF
associated with $\tilde A(\phi,\psi,\theta,\zeta)$.  The map is in
remarkably good agreement with the one in Fig.~\ref{fig:7}.

As a further test of accuracy, we re-calculated the mean force
using~(\ref{eq:tamd2}) and~(\ref{eq:meanforceapprox}) using a
different set of centers than those used in~(\ref{eq:radialbasis}).
Then we estimated the relative error between these mean forces and the
ones obtained by taking the negative gradient of the
reconstructed~$\tilde A(\phi,\psi,\theta,\zeta)$ using the original
set of centers and mean forces:
\begin{equation}
  \label{eq:relerror}
  \eps_k = \frac{|\nabla_{\!z}\tilde A(\zz^n_{k}) 
  + \ff^n_{\!k}|}{| \ff^n_{\!k}|}
\end{equation}
where $\zz^n_k$ are the new centers and $\ff^n_{\!k}$ are the mean
forces at these centers. The new centers were 20 points chosen at
random in the domains
$\phi,\psi\in[-180^\circ,180^\circ]$,~$\theta,\zeta\in[-70^\circ,70^\circ]$.

\begin{figure}[t]  
  \centerline{\includegraphics[width=3.25in]{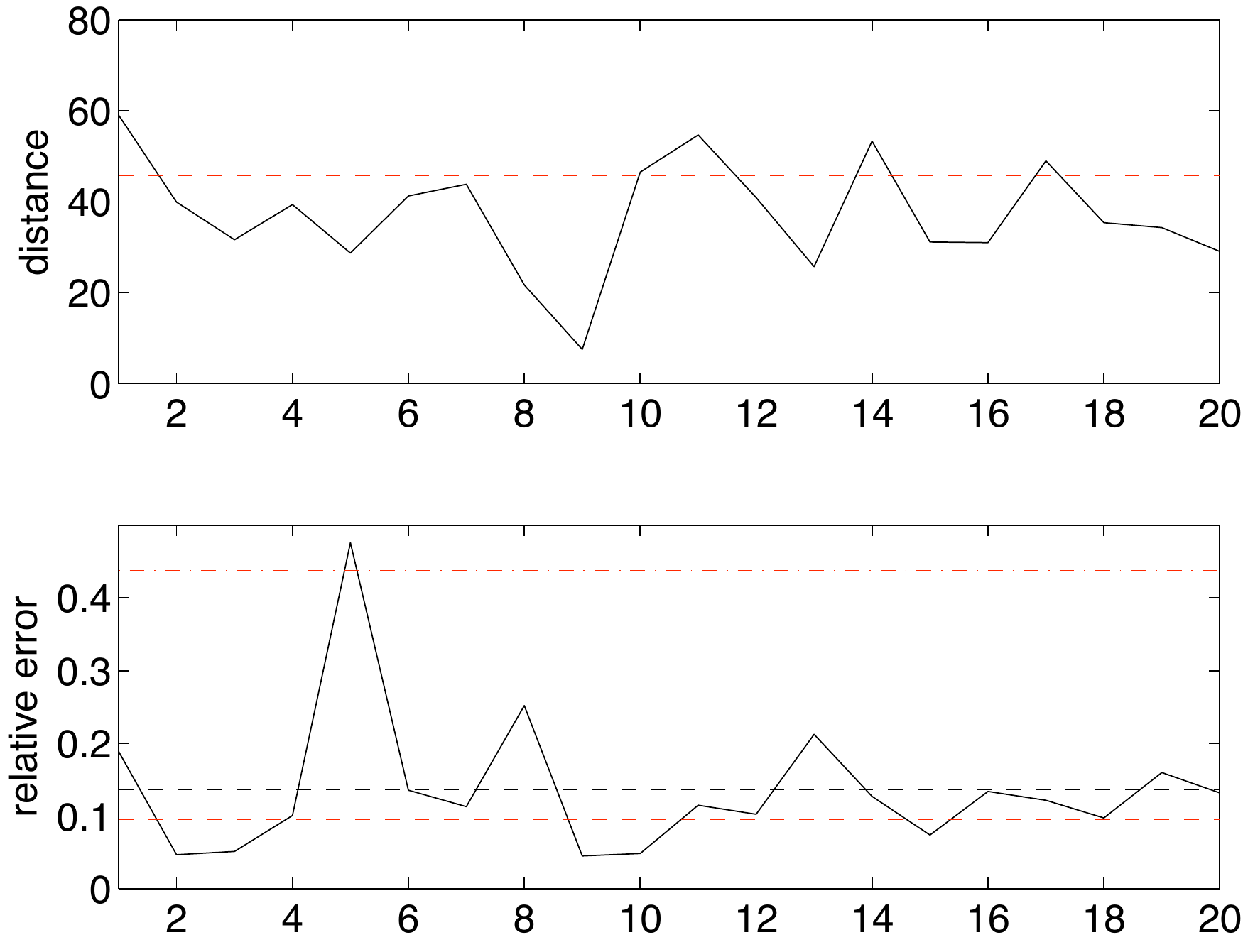}}
  \vspace{.5cm}
  \centerline{\includegraphics[width=3.25in]{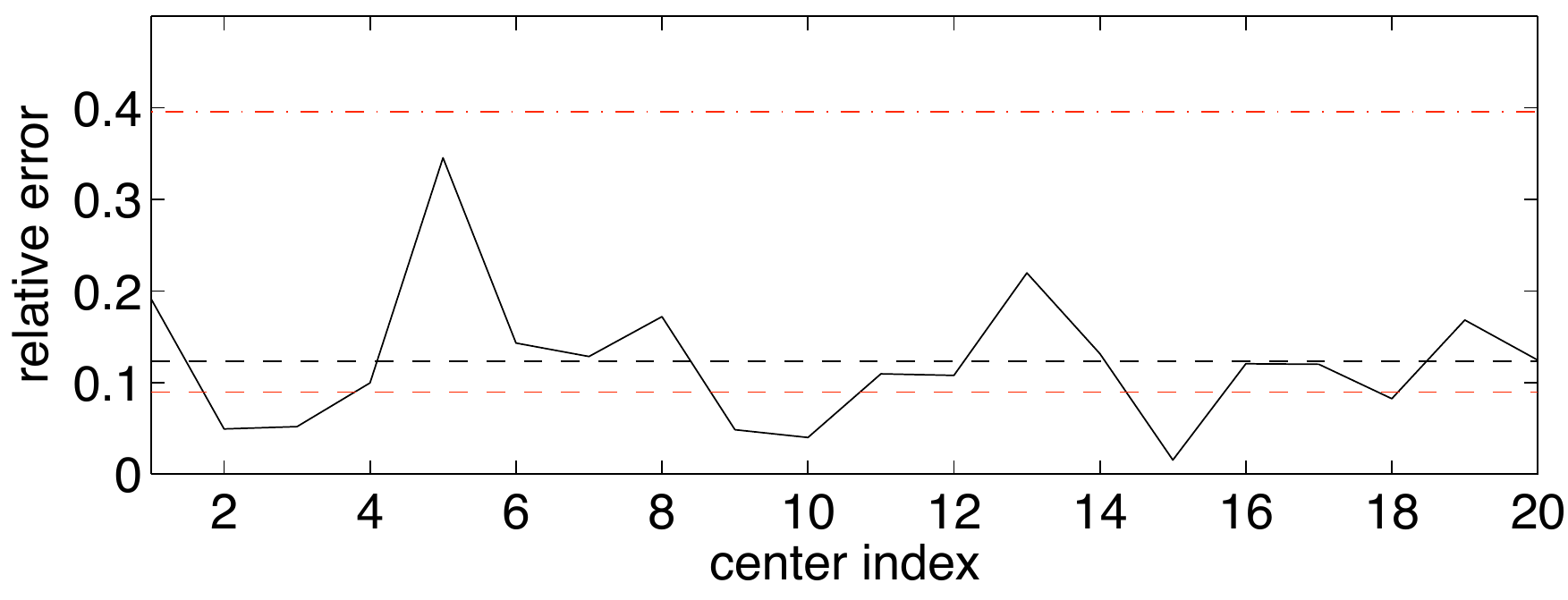}}
  \caption{% 
    Accuracy of the reconstruction of the free energy of AD in four
    angles. Top panel, distance of each of the random centers from the
    closest of the $200$ centers (black line), compared with the
    minimal distance between the $200$ centers (red line). Middle and
    lower panel, relative error $\varepsilon_k$ defined
    in~(\ref{eq:relerror}) for mean forces computed respectively from
    $50$ and $250$~ps restrained simulations: the error
    $\varepsilon_k$ (black solid line) and its mean value (black
    dashed line), compared with the mean value of the relative
    residual per center for the $200$ centers set (red dashed line)
    and its maximum value (red dashed-dotted line).}
  \label{fig:randtest}
\end{figure}

Fig.~\ref{fig:randtest}, top panel, shows, for each of these centers,
the distance from the closest of the $200$ centers (black line). Data
are compared to the minimal distance between the $200$ centers (red
dashed line). This result shows that, with $d=45.84^\circ$, these
centers fill properly the four dimensional domain in the sense that
every new center is always a distance about~$d$ to one of the original
centers.  Fig.~\ref{fig:randtest}, middle panel, shows the relative
error $\varepsilon_k$ for $k=1,\dots,20$ (black solid line), when
$T=50$~ps long restrained simulations are used to compute the mean
forces. The mean value of $\varepsilon_k$ (black dashed line) is
$0.14$, with standard deviation $0.09$ and maximum value $0.47$. For
comparison, the mean value of the relative residual per center (red
dashed line) is $0.09$, with standard deviation $0.06$ and maximum
value $0.44$ (red dashed-dotted line).  Fig.~\ref{fig:randtest},
bottom panel, shows $\varepsilon_k$ when $T=250$~ps long restrained
simulations are used to compute the mean forces. In this case, the
mean value of $\varepsilon_k$ (black dashed line) is $0.12$, with
standard deviation $0.07$ and its maximum value is $0.34$. For
comparison, the mean value of the relative residual per center (red
dashed line) is $0.09$, with standard deviation $0.06$ and maximum
value $0.39$ (red dashed-dotted line).

These results show that, in points away from the original centers, the
reconstructed free energy is as accurate as it is at the centers,
which is clearly the best we can hope for.

\section{Concluding remarks}
\label{sec:conclu}

In summary, we have proposed a method for the calculation of free
energies which is simple, accurate, and efficient. Unlike standard
histogram methods such as WHAM and metadynamics, the single-sweep
method uses the mean force computed at a set of centers to reconstruct
the free energy. This set of centers is determined using TAMD to
rapidly sweep through the important regions of the free energy, and
the mean forces at these centers are estimated in a standard way via
the computation of a conditional expectation using time-averaging
along restrained or constrained simulations. From these data, the free
energy $A(\zz)$ is then reconstructed globally by minimization of an
objective function to determine the coefficients in a radial-basis
function representation of $A(\zz)$. If convenient, this
reconstruction step can use data for the centers and the mean forces
obtained by other means than TAMD.

Compared with histogram methods and
metadynamics, the single-sweep technique combines several advantages:

\begin{itemize}
\item It does not require \textit{a~priori} knowledge of the free
  energy since it uses TAMD to find the important regions in the
  landscape automatically.
\item The most costly step of the calculation, namely the computation
  of the mean forces at the centers, can be straightforwardly
  distributed on different, independent, processors. 
%  where they are performed
%  independently.
\item The reconstruction step is variational, i.e. the optimal
  coefficients in the free energy representation are determined
  automatically, which limits the number of parameters to adjust
  beforehand.
\item The results can be easily monitored for convergence, and
  systematically improved if desired. In particular, new centers can
  be added on top of previous ones along the same TAMD trajectory to
  increase the accuracy without having to repeat the previous
  calculation.  
\item The method can be used in more than 2 dimensions and its
  computational complexity is the same regardless of the dimension.
\end{itemize}

We believe that these features make the single-sweep method appealing
to calculate the free energy of systems more complicated but also more
interesting than the ones studied in this paper.

\section*{Acknowledgments} 
We thank Giovanni Ciccotti and David Chandler for carefully reading
the manuscript; Weinan E for pointing out sparse grid methods which
prompted us to test our method in four dimensions; Ron Elber and
Anthony West for their help with the MOIL code; Sara Bonella, Simone
Meloni, Michele Monteferrante and Maddalena Venturoli 
for useful discussions; and finally,
Eric Darve for suggesting the test using~(\ref{eq:relerror}).  This
work was partially supported by NSF grants DMS02-09959 and
DMS02-39625, and by ONR grant N00014-04-1-0565.

\appendix

\section{Details of the MD simulations}
\label{sec:MD}

All MD simulations were performed with the MOIL code~\cite{moil}, and
the AMBER/OPLS \cite{ff} force field as implemented in the code. 
A starting structure for the AD molecule 
(CH$_3$-CO-NH-C$_\alpha$HCH$_3$-CO-NH-CH$_3$) was solvated in
a box of $252$ water molecules of volume $(20~\text{\AA})^3$.
Periodic boundary conditions were used. Van der Waals interactions
were truncated at $9$~\AA.  Electrostatic interactions were treated
with the Particle Mesh Ewald method~\cite{spme} with real space cutoff
$9$~\AA, a grid of $32^3$ points, and $4$-th order \textit{B}-splines
for the interpolation of the structure factor (in order to be in the
high accuracy range~\cite{spme}). The TIP3 model~\cite{tip3} was used
for the water molecules. Non-bonded interaction lists were updated
every $10$ steps.  All chemical bonds in the system were kept fixed
with the SHAKE algorithm~\cite{shake}.  Amide planes were restrained
to be always in \textit{trans} configuration. The Velocity Verlet
algorithm was used for the dynamics of the Cartesian variables with
time-step $1$~fs, and all velocities were scaled at every step to keep
the temperature at $300$~K. In order to obtain the initial configuration 
for the temperature
accelerated MD simulation (TAMD), the system was first equilibrated
for $100$~ps by keeping the solute molecule fixed (i.e. by zeroing
forces and velocities of its atoms) and by assigning to all water
atoms at every step velocities sampled from a Maxwell distribution at
$300$~K.  Then, the whole system was simulated for $400$~ps for
equilibration. The torsion angles used in the simulations are defined by the
quadruplets of atoms (C,N,C$_\alpha$,C) and (N,C$_\alpha$,C,N) for
$\phi$ and $\psi$, and (O,C,N,C$_\alpha$) and (C$_\alpha$,C,N,H) for
$\theta$ and $\zeta$.  In the TAMD simulation, the equations of motion
of the collective variables were integrated with the forward Euler
scheme with time-step $1$~fs, $\gamma=0.5$~kcal$\times$ps/mol/rad$^2$
for $\phi$ and $\psi$ and $\gamma=1$~kcal$\times$ps/mol/rad$^2$ for
$\theta$ and $\zeta$.  The force constant for the restraint potential
was $\kappa=100$~kcal/mol/rad$^2$, and the effective temperature such
that $\bar \beta^{-1} =9.5$~kcal/mol. The Cartesian coordinates of water 
and AD atoms were saved during the
TAMD simulation.  In this way, for every center $\zz_k$ deposited
along the trajectory in collective variables space, there is a
corresponding configuration $\xx_k$ of the system in Cartesian space
such that $\theta(\xx_k)\approx \zz_k$. We used these configurations
as initial conditions for the restrained simulations at the centers.
Data for the mean force calculations were accumulated after further
relaxation of the system for $5$~ps.


\begin{thebibliography}{99}
\bibitem{rem1} A term $m_z \ddot \zz$ is typically added
  to~(\ref{eq:zzmeta}) and the thermostat can be different, but this
  is not essential for our discussion and does not change our
  conclusions.
\end{thebibliography}
\end{document}